\documentclass[aps,pra,twocolumn,notitlepage,groupedaddress, longbibliography]{revtex4-1}  
\setcitestyle{square}
\usepackage{graphicx}  
\usepackage{float}
\usepackage{dcolumn}   
\usepackage{bm}        
\usepackage{amssymb}   
\usepackage{amsmath}
\usepackage{isomath}
\usepackage[usenames, dvipsnames]{color}
\usepackage{soul}
\usepackage{physics}
\usepackage[normalem]{ulem}
\usepackage{xpatch}
\usepackage{IEEEtrantools}

\usepackage{xcolor}
\usepackage{hyperref}
\hypersetup{
     colorlinks   = false,
     citecolor    = blue
}

\newcommand{\w}{\mathrm{\Omega}}
\newcommand{\D}{\mathrm{\Delta}}
  
\DeclareMathOperator{\sgn}{sgn}
  
\begin{document}

\widetext
\title{Self-induced topological transition in phononic crystals \\ by nonlinearity management }

\author{Rajesh Chaunsali}
\email{rajeshcuw@gmail.com}
\affiliation{LAUM, CNRS, Le Mans Universit\'{e}, Avenue Olivier Messiaen, 72085 Le Mans, France}

\author{Georgios Theocharis}
\email{georgiostheocharis@gmail.com}
\affiliation{LAUM, CNRS, Le Mans Universit\'{e}, Avenue Olivier Messiaen, 72085 Le Mans, France}

\date{\today}

\begin{abstract}
A new design paradigm of topology has recently emerged to manipulate the flow of phonons. At its heart lies a topological transition to a nontrivial state with exotic properties. This framework has been limited to linear lattice dynamics so far. Here we show a topological transition in a nonlinear regime and its implication in emerging nonlinear solutions. We employ nonlinearity management such that the system consists of masses connected with two types of nonlinear springs, ``stiffening" and ``softening" types, alternating along the length. We show, analytically and numerically, that the lattice makes a topological transition simply by changing the excitation amplitude and invoking nonlinear dynamics. Consequently, we witness the emergence of a new family of finite-frequency edge modes, not observed in linear phononic systems. We also report the existence of kink solitons at the topological transition point. These correspond to heteroclinic orbits that form a closed curve in the phase portrait separating the two topologically-distinct regimes. These findings suggest that nonlinearity can be used as a strategic tuning knob to alter topological characteristics of phononic crystals. These also provide fresh perspectives towards understanding a new family of nonlinear solutions in light of topology.

\end{abstract}
\maketitle

\section{Introduction}  \label{Section1}
The advent of topological insulators (TIs) in condensed matter physics has popularized a tool to characterize material dispersion~\citep{Hasan2010}. This tool is the topology of energy bands. It offers a powerful framework of the bulk-boundary correspondence, in which one can characterize a bulk (infinite) material topologically, and predict the response at the boundary of its finite counterpart. At a physical level, this explains why TIs allow chiral and robust currents on their boundaries but do not support any current in their bulk. Recently, this framework has been extended beyond electronic systems and has been applied to classical systems in photonics~\citep{Lu2014} and phononics \citep{Susstrunk2016}. For example, acoustic and elastic topological structures show various possibilities of manipulating wave flow~\citep{Ma2019}. These have the potential to provide novel solutions for applications, such as noise mitigation, vibration isolation, sensing, and energy harvesting. At the same time, these structures act as table-top setups for investigating topological physics at a fundamental level.

The study of band topology in phononic lattices has been largely restricted to linear dynamics so far. It is not clear how this framework could be relevant for nonlinear systems as well, where wave dispersion becomes amplitude dependent, at least in weakly nonlinear regimes. Previous studies have shown the excellent potential of nonlinear phononic crystals manipulating vibrations in general (see ~\citep{Theocharis2013, Hussein2014} and references therein). Especially appealing are elastic systems in which a tremendous degree of nonlinearity management could be achieved through material and geometric parameters, for example, using contacts~\citep{Porter2015}, LEGO blocks~\citep{Deng2018}, origami folding~\citep{Yasuda2019}, tensegrity structures~\citep{Fraternali2015}, or architected soft media~\citep{Raney2016}. 
All these structures and other proposed flexible mechanical metamaterials~\citep{Bertoldi2017} can thus be excellent model candidates for the fundamental understanding of the interplay of nonlinearity and topology in mechanics. 
At the same time, the findings can be experimentally verified, and also supplement other active areas, such as nonlinear photonics, where exciting theoretical advances have been made recently~\cite{Lumer2013, Ablowitz2014, Leykam2016, Lumer2016, Hadad2016, Solnyshkov2017, Dobrykh2018, Poddubny2018, Savelev2018}, but with limited experimental success~\citep{Dobrykh2018, Kruk2019}.

One may ask why is the study of nonlinearity relevant in topological systems after all. In this quest, the recent studies in both photonics and phononics have shown two broad pathways. The first is to assess the effect of nonlinearity, inevitable in most systems at large operating amplitudes, on the topologically-robust edge states present in a linear system. For example, one can investigate the change in frequency and stability of edge modes~\cite{Lumer2016, Dobrykh2018, Pal2018, Kruk2019}, or reveal the formation of topologically-robust edge solitons~\citep{Ablowitz2014, Leykam2016, Snee2019}. 
The second pathway is to explore strategically-designed nonlinear lattices, in which nonlinear solutions are intrinsically linked to topological nature of the lattice such that unique gap solitons~\citep{Lumer2013, Solnyshkov2017}, and ``self-induced'' edge solitons~\citep{Leykam2016} and domain walls~\citep{Chen2014, Hadad2017, Poddubny2018} appear in the lattice. 
We focus on the latter category since it provides novel ways to \textit{complement} topological effects by using nonlinearity management and at the same provide a fundamentally new way to interpret emerging nonlinear solutions in light of topology.

In phononics, one of the first such works~\citep{Chen2014} studied the dynamics of zero-frequency floppy modes in a 1D mechanical chain, the exact mapping of the celebrated SSH model~\citep{Su1979, Kane2014}. It was shown that these modes can travel from one end to the other in the form of a soliton along the chain, and the soliton is interpreted as a moving domain wall in a topological sense. 
However, when it comes to topological band theory for finite (nonzero)-frequency modes~\citep{Susstrunk2016}---relevant for vibration studies---it is not clear how emergent nonlinear solutions, such as ``self-induced'' solitons and domain wall, are connected with band topology of phonons. 

To address this question, here, we consider one of the most fundamental mechanisms in topological band theory---the band inversion. We take a phononic crystal, a chain made of masses and two alternating springs and ask the question: Can we engineer a nonlinear phononic lattice that shows a band inversion, a topological transition, and support emergent edge modes solely based on the amplitude of vibration excitation [Fig.~\ref{fig1}(a)]?  
The model is inspired from the SSH chain~\citep{Su1979}, and the difference lies in the fact that spring stiffnesses contribute to the diagonal elements of the bulk Hamiltonian, and therefore, the bandgap is centered at a finite frequency~\citep{Chaunsali2017}. 
If this ``self-induced" topological transition could be engineered, we expect to observe a family of emerging nonlinear solutions with the increase in excitation amplitude and thus establish any connection they might have with topological band theory. By introducing nonlinearity in \textit{all} the springs, different from the previous studies~\citep{Hadad2016, Hadad2018a}, we extensively rely on phase portrait analysis to unveil a rich nonlinear dynamics in the system. In this process, we discover stationary kink soliton solutions, and those can be interpreted as a ``self-induced'' domain wall separating two topologically-distinct regions within the purview of an ``effective'' topological band theory of our nonlinear lattice. Moreover, we study how nonlinear effects lead to unique edge solutions in the system depending on excitation amplitude and how this fact could be used for designing tunable topological systems.

We organize this manuscript as follows: In Section~\ref{Section2}, we propose a scheme of nonlinearity management, using a combination of ``stiffening" and ``softening" type of nonlinear springs with cubic nonlinearity. In Section~\ref{Section3}, we describe amplitude-dependent dispersion properties and general solutions, and show the mechanism of a band inversion and thus a topological transition. In Section~\ref{Section4}, we employ analytical techniques and use phase portrait to reveal emerging nonlinear solutions in the bulk, such as nonlinear evanescent modes and kink solitons. We establish a connection between  these nonlinear solutions and band topology. In Section~\ref{Section5}, we use numerical techniques to verify their existence for finite structures. In Section~\ref{Section6}, for experimental feasibility, we also extend these results for a ``saturating" type of nonlinearity and show how this system can support an in situ emergence of an edge mode due to a topological transition. In Section~\ref{Section7}, we conclude our findings and provide fresh perspectives towards extending this work to different topological systems and also realizing in experiments.

\section{System details} \label{Section2}
 \begin{figure}[t]
\centering
\includegraphics[width=3.3in]{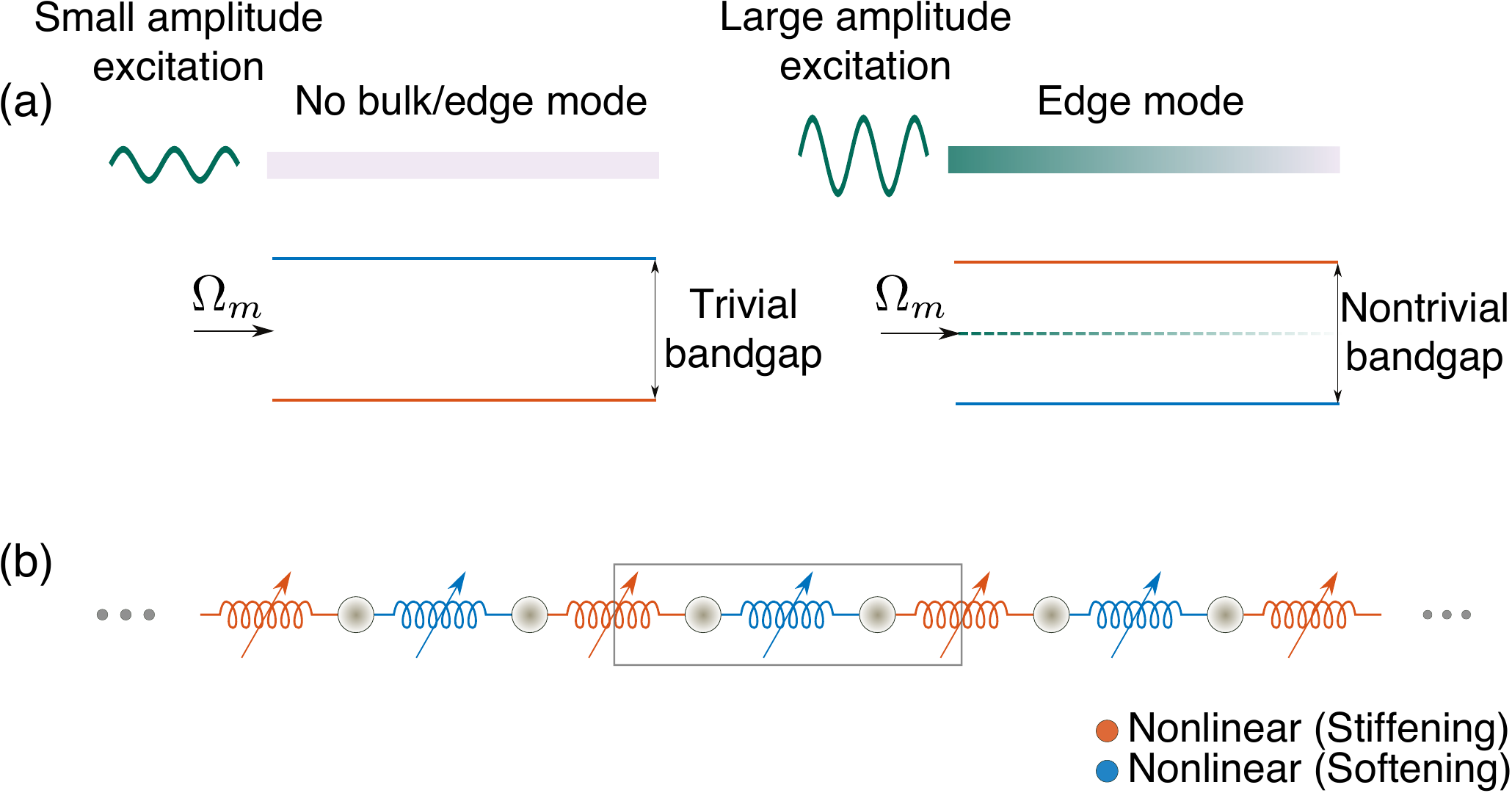}
\caption{[Color online](a) A schematic of the ``self-induced'' topological transition in the system. For small amplitude excitation, the system has a trivial bandgap, with no edge or bulk mode at the mid-gap frequency $\mathrm{\Omega}_m$. However, for large amplitude, nonlinearity kicks in and forces the system to a nontrivial state. This leads to the emergence of an edge mode. (b) A chain of point masses and nonlinear springs (stiffening in red and softening in blue) is designed to induce the aforementioned effect. The box highlights the unit cell.}
\label{fig1}
\end{figure} 

Figure~\ref{fig1}(b) shows the system configuration, which is a chain of point masses connected with nonlinear springs. We use ``stiffening'' and ``softening'' type of springs alternatively along the chain. The unit cell consists of two masses. For the first section of this paper, we take the following force-deformation profiles with cubic nonlinearity for two types of springs:
\begin{equation}
\left.
\begin{IEEEeqnarraybox}[
\IEEEeqnarraystrutmode
\IEEEeqnarraystrutsizeadd{2pt}
{2pt}
][c]{rCl}
\text{F}( \mathrm{\Delta} x)&=&(1-\gamma) k \mathrm{\Delta} x + k_3 ( \mathrm{\Delta} x)^3  \; \;  \textrm{(stiffening)}, \\
\text{F}( \mathrm{\Delta} x)&=&(1+\gamma) k   \mathrm{\Delta} x - k_3 ( \mathrm{\Delta} x)^3  \; \; \textrm{(softening)},
\end{IEEEeqnarraybox}
\, \right\}
\label{eq:cubic}
\end{equation}

\noindent where $ \mathrm{\Delta} x$ and $\mathrm{F}$ are spring deformation and force, respectively; $k>0$ and $k_3>0$ are the stiffness coefficients; $\gamma>0$ denotes the mismatch in stiffness for small amplitude (linear dynamics) and this determines the width of the initial bandgap. We can thus write equations of motion neglecting any dissipation as
 
\begin{IEEEeqnarray}{rCl}
m \frac{\mathrm{d}^2 x_{1,n}}{\mathrm{d}t^2} &=&\text{F}(x_{2,n-1}-x_{1,n})-\text{F}(x_{1,n}-x_{2,n}) \nonumber \\ 
&=&(1-\gamma)k (x_{2,n-1}-x_{1,n})+k_3 (x_{2,n-1}-x_{1,n})^3  \nonumber \\
&&-\> (1+\gamma)k (x_{1,n}-x_{2,n})+k_3 (x_{1,n}-x_{2,n})^3,  \nonumber \\
m \frac{\mathrm{d}^2 x_{2,n}}{\mathrm{d}t^2} &=& \text{F}(x_{1,n}-x_{2,n})- \text{F}(x_{2,n}-x_{1,n+1}) \nonumber \\ 
&=&(1+\gamma)k (x_{1,n}-x_{2,n})-k_3 (x_{1,n}-x_{2,n})^3 \nonumber \\
&&-\> (1-\gamma)k (x_{2,n}-x_{1,n+1})-k_3 (x_{2,n}-x_{1,n+1})^3, \nonumber 
\end{IEEEeqnarray}

\noindent where $x_{1,n}$ and $x_{2,n}$ denote displacements of two particles inside the $n$th unit cell. 
Let the unit-cell length be $a$ and $ \mathrm{\Gamma} = a^2 k_3/k $. Therefore, we can non-dimensionalize equations of motion by defining new variables $\tau=t \sqrt{k/m}$ and  $\xi=x/a$, and write 
\begin{equation}
\left.
\begin{IEEEeqnarraybox}[
\IEEEeqnarraystrutmode
\IEEEeqnarraystrutsizeadd{2pt}
{2pt}
][c]{rCl}
\ddot{\xi}_{1,n} &=& (1-\gamma) (\xi_{2,n-1}-\xi_{1,n})+\mathrm{\Gamma} (\xi_{2,n-1}-\xi_{1,n})^3  \\ 
&& -\>(1+\gamma) (\xi_{1,n}-\xi_{2,n})+\mathrm{\Gamma} (\xi_{1,n}-\xi_{2,n})^3,   \\
 \ddot{\xi}_{2,n}&=&(1+\gamma) (\xi_{1,n}-\xi_{2,n})- \mathrm{\Gamma} (\xi_{1,n}-\xi_{2,n})^3  \\ 
&&- \>(1-\gamma) (\xi_{2,n}-\xi_{1,n+1})- \mathrm{\Gamma} (\xi_{2,n}-\xi_{1,n+1})^3.
\end{IEEEeqnarraybox}
\, \right\}
\label{eq:eom}
\end{equation}

\noindent For the rest of the paper, we stick to this non-dimensionalized framework, where $\mathrm{\Gamma}$, $\tau$, $\xi$, and $\mathrm{\Delta} {\xi}$ denote nonlinearity parameter, time, displacement, and strain, respectively. 

\section{Bandgap solutions, and topological transition with amplitude}  \label{Section3}
 \begin{figure}[t]
\centering
\includegraphics[width=3.3in]{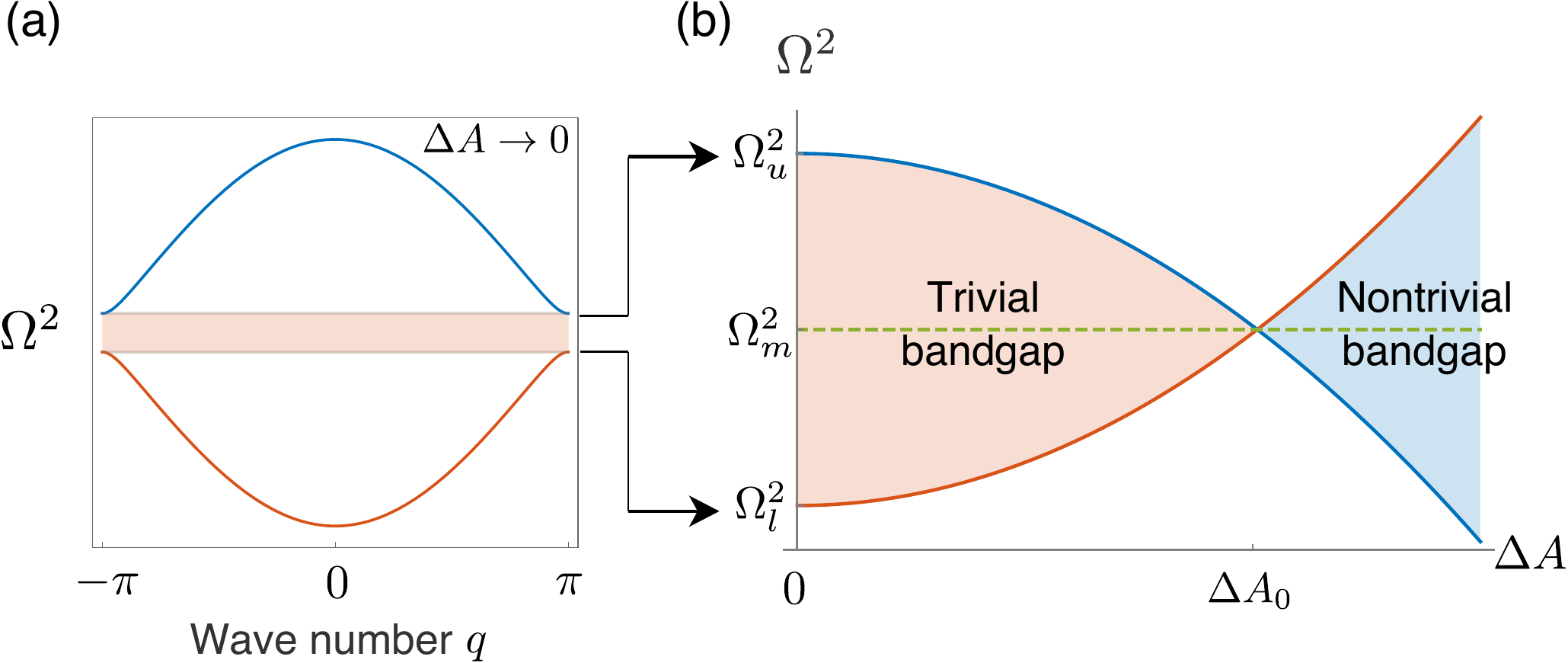}
\caption{[Color online] (a) Dispersion relation of the unit cell with small strain amplitude $\mathrm{\Delta}A$, a linear regime. It highlights a bandgap between acoustical (lower) and optical (upper) branches. (b) Increase in strain amplitude $\mathrm{\Delta}A$ changes the bandgap, by  closing it first and then opening again to a nontrivial state.}
\label{fig2}
\end{figure} 

\subsection{Bandgap solutions}
In the linear limit, the system represents a periodic chain consisting of two alternating springs with stiffness $1-\gamma$ and $1+\gamma$. The dispersion relation, as shown in Fig.~\ref{fig2}(a), consists of two branches, the acoustical (lower) with $\w_l^2 = 2(1-\gamma)$ at $q= \pm \pi$, and the optical (upper) with  $\w_u^2 = 2(1+\gamma)$ at $q= \pm \pi$. Thus the system has a bandgap of width  $ \w_u^2-\w_l^2$. It is straightforward to show that the eigenmode corresponds to $\w_l^2$ is given by $(\xi_{1,n}, \xi_{2,n})=(A ,A) \exp(i  q n)=(-1)^n(A ,A)$, where $A$ represents the amplitude of oscillation. Physically, this means that the two particles inside the unit cell oscillate \textit{in phase}. Similarly, the eigenmode at $\w_u^2$ is given by $(\xi_{1,n}, \xi_{2,n})=(A ,-A)\exp(i  q n)=(-1)^n(A ,-A)$, representing \textit{out-of-phase} motion of particles inside the unit cell.

In this work, we focus on the weakly nonlinear wave solutions, i.e., $ \mathrm{\Gamma}=O(\epsilon)$ inside a small bandgap with $\gamma= O(\epsilon)$. Therefore we can look for slowly-varying solutions around frequency $\w$ with the following ansatz:


\begin{equation}
\left.
\begin{IEEEeqnarraybox}[
\IEEEeqnarraystrutmode
\IEEEeqnarraystrutsizeadd{2pt}
{6pt}
][c]{rCl}
\xi_{1,n} =\frac{ (-1)^n }{2} \left [u(n,t) \exp(i \mathrm{\Omega} \tau) + u^*(n,t) \exp(-i \mathrm{\Omega} \tau) \right], \\
\xi_{2,n}= \frac{ (-1)^n }{2} \left [v(n,t) \exp(i \mathrm{\Omega} \tau) + v^*(n,t) \exp(-i \mathrm{\Omega} \tau) \right], 
\end{IEEEeqnarraybox}
\, \right\}
\label{ansatz}
\end{equation}
\noindent where * indicates the complex conjugation and $\w \in [\w_l, \w_u]$. We substitute the ansatz in Eq.~\eqref{eq:eom} and make continuum approximation. To this end, we choose $n=z$, $u(n-1,t)=u(z,t)-{\partial u(z,t)}/{\partial z}$, $u(n+1,t)=u(z,t)+{\partial u(z,t)}/{\partial z}$, $v(n-1,t)=v(z,t)-{\partial v(z,t)}/{\partial z}$, and $v(n+1,t)=v(z,t)+{\partial v(z,t)}/{\partial z}$, where the derivatives are small of $O(\epsilon)$. We ignore higher harmonics and collect the terms of $O(\epsilon)$ to obtain the following PDEs:

\begin{equation}
\left.
\begin{IEEEeqnarraybox}[
\IEEEeqnarraystrutmode
\IEEEeqnarraystrutsizeadd{2pt}
{6pt}
][c]{rCl}
- 4 i \w \frac{\partial u(z,t)}{\partial t} &=& -2  \frac{\partial v(z,t)}{\partial z}  \\
&&+\> 2 \left(2-\w^2 \right) u(z,t)  - 4 \gamma v(z,t)  \\
&&+\>  3 \mathrm{\Gamma} \left [2 \abs{u(z,t)}^2 + \abs{v(z,t)}^2 \right ]v(z,t) \\
&&+\> 3 \mathrm{\Gamma} {u(z,t)}^2 v^*(z,t),  \\
- 4 i \w \frac{\partial v(z,t)}{\partial t} &=& 2  \frac{\partial u(z,t)}{\partial z}  \\
&&+\> 2 \left(2-\w^2 \right) v(z,t)  - 4 \gamma u(z,t)  \\
&&+\>  3 \mathrm{\Gamma} \left [\abs{u(z,t)}^2 + 2\abs{v(z,t)}^2 \right ]u(z,t) \\
&&+\> 3 \mathrm{\Gamma} {v(z,t)}^2 u^*(z,t).  
\end{IEEEeqnarraybox}
\, \right\}
\label{eq:dirac}
\end{equation}

For steady-state solutions, we can simply substitute $u(z,t)=u(z)=u$ and $v(z,t)=v(z)=v$ in Eq.~\eqref{eq:dirac} and deduce the following ODEs:
\begin{equation}
\left.
\begin{IEEEeqnarraybox}[
\IEEEeqnarraystrutmode
\IEEEeqnarraystrutsizeadd{7pt}
{7pt}
][c]{rCl}
u'&=& - (2-\w^2)v + 2 \gamma u - \frac{3}{2}\mathrm{\Gamma}(u^3+3u v^2), \\
v'&=&  (2-\w^2)u - 2 \gamma v + \frac{3}{2}\mathrm{\Gamma}(v^3+3u^2 v),
\end{IEEEeqnarraybox}
\, \right\}
\label{eq:ode}
\end{equation}

 \noindent where $'$ denotes first derivative w.r.t. $z$, and $u$ and $v$ are real quantities.

\subsection{Bandgap closure with amplitude}
We calculate the nonlinear continuation of the lower and upper band-edge modes by substituting $u=v=A$ for $\w=\w_l$,  and $u=-v=A$ for $\w=\w_u$  in Eq.~\eqref{eq:ode}. These represent steady-state and spatially periodic solutions with amplitude $A$. 
We use the fact that for these modes, \textit{strain} amplitudes ($\D A$) are $2A$ and 0 for the two springs inside the unit cell, therefore

\begin{equation}
\left.
\begin{IEEEeqnarraybox}[
\IEEEeqnarraystrutmode
\IEEEeqnarraystrutsizeadd{7pt}
{7pt}
][c]{rCl}
\mathrm{\Omega}_l^2(\D A)=2 (1-\gamma)+ \frac{3}{2}  \mathrm{\Gamma} (\D A)^2,  \\
\mathrm{\Omega}_u^2(\D A)=2 (1+\gamma)-  \frac{3}{2}   \mathrm{\Gamma} (\D A)^2.
\end{IEEEeqnarraybox}
\, \right\}
\label{eq:freqNL}
\end{equation}

\noindent We plot the band-edge continuation in Fig.~\ref{fig2}(b). We see that the bandgap first closes at $\mathrm{\Delta}A_{0}=2\sqrt{\gamma /3 \mathrm{\Gamma}}$ and then opens again, with band-edge frequencies flipped---indicating a band inversion. Note that such a band inversion could be achieved by other nonlinearity management, for example, by taking linear and nonlinear springs alternating along the chain as was done in photonics~\citep{Hadad2016}. However, by using a combination of stiffening and softening type of nonlinear springs in this study, we not only achieve a band inversion, but also we make sure that the mid-gap solutions (point of interest in 1D topological systems) occur at the \textit{same} frequency, $\mathrm{\Omega}_m^2=[\mathrm{\Omega}_l^2 (\mathrm{\Delta}A)+\mathrm{\Omega}_u^2 (\mathrm{\Delta}A)]/2=2$, for different amplitude excitation.

\subsection{Topological transition}
One can characterize the band topology of a one-dimensional linear system using the so-called Zak phase~\citep{Zak1989, Delplace2011}.  This is the geometric phase accumulated by the Bloch wave vector when transported adiabatically along a frequency band in the first Brillouin zone. For a linear dimer system with inversion symmetry, it is either 0 or $\pi$. For the unit cell shown in Fig.~\ref{fig1}(b) but made of linear springs, one can easily verify that it would have a nonzero Zak phase $\pi$ when the stiffness of red spring is \textit{larger} than the stiffness of blue spring. In other words, a finite-frequency SSH model would have a nonzero Zak phase when intercell stiffness ($\text{K}_{\textrm{inter}}$) would be larger than intracell stiffness ($\text{K}_{\textrm{intra}}$), such that~\citep{Chaunsali2017}

\begin{equation}
\mathcal{Z}= \frac{\pi}{2} \big [1 + \sgn \big (\text{K}_{\textrm{inter}} -\text{K}_{\textrm{intra}} \big ) \big ], \nonumber
\end{equation}

\noindent where $\sgn(.)$ denotes the sign function. Now the question is: How could we utilize this formulation of the Zak phase in a nonlinear system? The question is highly nontrivial since Bloch vectors become amplitude dependent. This implies that the geometric phase accumulated by the Bloch vector along a frequency band could then vary depending on how much amplitude excitation we give at each frequency, and this fact would defeat the basic definition of a bulk invariant. Therefore, we argue that such a nonlinear system would not have a bulk invariant in the sense of a linear system. However, for a weakly nonlinear system as ours, where the Bloch theorem is still valid, we map our nonlinear system to an \textit{effective} linear system at the unit-cell level. For nonlinear excitations inside the small bandgap, we thus write effective stiffnesses for the two type of nonlinear springs (see Appendix A for more discussion)

\begin{equation}
\left.
\begin{IEEEeqnarraybox}[
\IEEEeqnarraystrutmode
\IEEEeqnarraystrutsizeadd{7pt}
{7pt}
][c]{rCl}
\text{K}_{\textrm{stiffening}}(\mathrm{\Delta}A) &=& \frac{\mathrm{\Omega}_l^2(\D A)}{2} = (1-\gamma) +\frac{3}{4}  \mathrm{\Gamma} (\mathrm{\Delta}A)^2, \\
\text{K}_{\textrm{softening}}(\mathrm{\Delta}A) &=& \frac{\mathrm{\Omega}_u^2(\D A)}{2} = (1+\gamma) -\frac{3}{4}  \mathrm{\Gamma} (\mathrm{\Delta}A)^2.
\end{IEEEeqnarraybox}
\, \right\}
\label{eq:es}
\end{equation}

\noindent We thus characterize this mapped linear system by  
\begin{equation}
\mathcal{Z} (\mathrm{\Delta}A) = \frac{\pi}{2} \big [1 + \sgn \big (\text{K}_{\textrm{stiffening}}(\mathrm{\Delta}A) -\text{K}_{\textrm{softening}}(\mathrm{\Delta}A) \big ) \big ]. 
\label{eq:zak}
\end{equation}

\noindent By using the expressions in Eq.~\eqref{eq:es}, our mapped system, therefore, makes a transition from a topologically-trivial state ($\mathcal{Z}=0$) to a topologically-nontrivial state ($\mathcal{Z}=\pi$) when strain amplitude crosses $\mathrm{\Delta}A_{0}$ and bands are inverted. 
One may, however, notice another assumption in this approach, and that is to assume the equal strain amplitude $\mathrm{\Delta}A$ in all the springs. To explain the existence of any edge or soliton solution using this approach is not straightforward since such non-uniform spatial solutions in a nonlinear system as ours, can change effective stiffness, locally, along the chain and the system no longer remains periodic from the effective stiffness point of view. Unit-cell description, fundamental in characterizing global topology, thus breaks down. Then the question is whether this simplified description of topology is of any use in the current case? The answer is: Yes. 
We will show in the sections below that, we do observe soliton and edge solutions in the system, whose presence could be explained by an amplitude-dependent topological transition if topological characterization is done \textit{locally} by calculating local effective stiffness.

\section{Phase portrait and emergence of soliton and evanescent solutions}  \label{Section4}
 \begin{figure*}[t]
\centering
\includegraphics[width=6.3in]{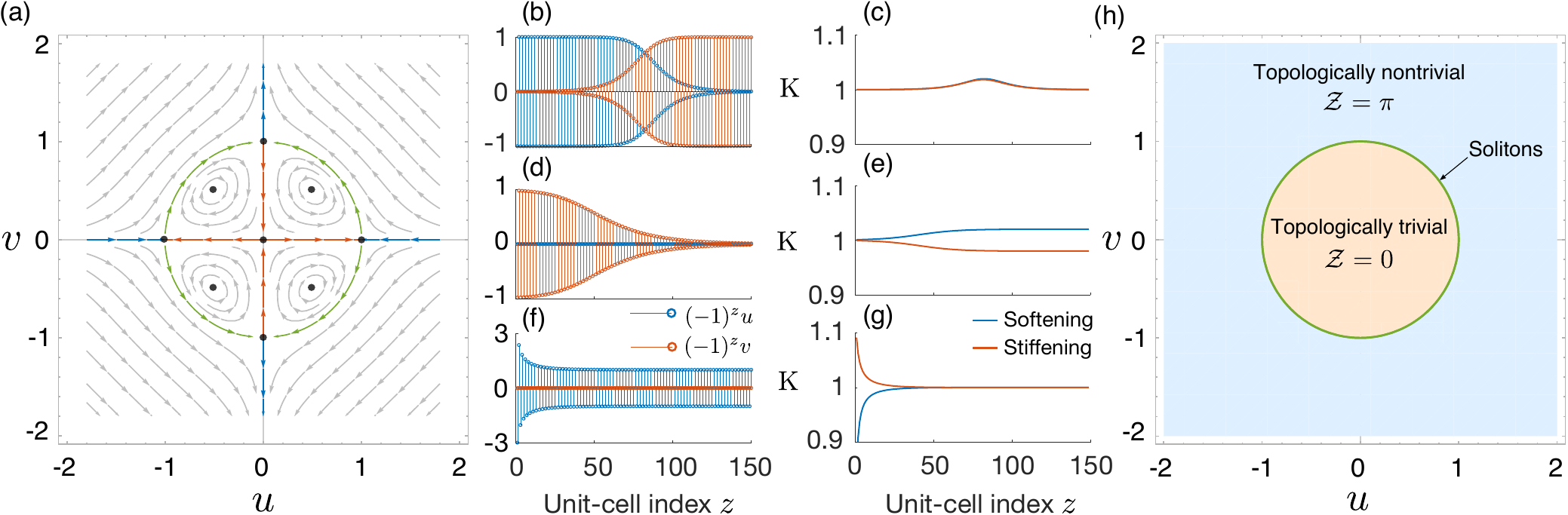}
\caption{[Color online] (a) Phase portrait of the \textit{bulk} showing steady-state solutions at the mid-gap frequency. It highlights 9 fixed points in black dots. The arrows in red, blue, and green indicate the presence of multiple evanescent and soliton solutions. 
(b) Soliton profile that corresponds to the heteroclinic orbits in green. 
(c) Effective stiffness for all the springs along the chain, calculated for the aforementioned mode profile. This indicates the closure of ``local bandgap''.
(d)-(e) For small amplitude regimes, a profile and its effective stiffness, respectively, extracted from one of the evanescent solutions shown in \textit{red} in the phase portrait. Note the amplitude forms a plateau towards the left edge.
(f)-(g) For large amplitude regimes, a profile and its effective stiffness, respectively, extracted from one of the evanescent solutions shown in \textit{blue} in the phase portrait. Note the amplitude forms a plateau towards the right edge.
(h) A schematic to show topologically distinct regions in that phase portrait that are separated by the circular trajectory of solitons.
}
\label{fig3}
\end{figure*} 

\subsection{Mid-gap steady-state solutions}
From now onward, we focus on the solutions at the mid-gap frequency $\mathrm{\Omega}_m^2=2$. Therefore, Eq.~\eqref{eq:ode} reduces to

\begin{equation}
\left.
\begin{IEEEeqnarraybox}[
\IEEEeqnarraystrutmode
\IEEEeqnarraystrutsizeadd{7pt}
{7pt}
][c]{rCl}
u'&=&2 \gamma u - \frac{3}{2}\mathrm{\Gamma}(u^3+3u v^2), \\
v'&=&- 2 \gamma v + \frac{3}{2}\mathrm{\Gamma}(v^3+3u^2 v).
\end{IEEEeqnarraybox}
\, \right\}
\label{eq:portrait}
\end{equation}

 \noindent  These describe the dynamics of a Hamiltonian system with a conserved energy
\begin{equation}
E =2 \gamma u v \left [1- \frac{3\mathrm{\Gamma}}{4 \gamma}(u^2+v^2) \right],
\end{equation}
\noindent where $u$ and $v$ may be considered as generalized coordinate and momentum, respectively. Using these equations, we now plot the phase portrait in Fig.~\ref{fig3}(a) with $4 \gamma = 3 \mathrm{\Gamma}= 0.08$ and look for special trajectories. 

We observe 9 fixed points ($u'=v'=0$) highlighted by black dots at $(u,v)=(0,0)$, $(\pm 2 \sqrt{\gamma / 3\mathrm{\Gamma}}, 0)$, $(0, \pm 2 \sqrt{\gamma/ 3\mathrm{\Gamma}})$, and $(\pm \sqrt{\gamma/3\mathrm{\Gamma}}, \pm \sqrt{\gamma/3 \mathrm{\Gamma}})$. Four of them, the ones at $(\pm 2 \sqrt{\gamma / 3\mathrm{\Gamma}}, 0)$, $(0, \pm 2 \sqrt{\gamma/ 3\mathrm{\Gamma}})$, are connected by heteroclinic orbits forming a circular curve in the phase portrait. Notice that the strain amplitude at these fixed points is exactly the same as $\D A_0$ shown in Fig.~\ref{fig2}(b). 
As we will see next, this circle separates the phase portrait into two topologically distinct regions.
For the choice of the unit cell, highlighted in Fig.1 (b) by the box, the region inside (outside) the circle corresponds to a topologically trivial (nontrivial) bandgap. Below we study in detail solutions corresponding to the two topologically distinct regimes as well as solutions at the topological transition on the circle.

\subsection{Topological transition point: Bulk soliton solutions}
In Fig.~\ref{fig3}(b), we plot the solution corresponding to the circular trajectory (in green) in the first quadrant. These are soliton solutions and represent two kinks of opposite polarities. On the left end of the chain, the unit cell has only the first mass moving (predominately $u$ component) while towards the right its polarity changes and only the second mass vibrates (predominantly $v$ component). Interesting enough, a similar soliton solution has been observed experimentally~\citep{Denardo1992} and studied analytically in some earlier works~\citep{Kivshar1992, Huang1993}, but in different types of nonlinear lattices.
We calculate the analytical expression of the soliton profile by finding that the soliton trajectories correspond to $E=0$ and  thus would have the locus given by
\begin{equation}
u^2+v^2= \frac{4 \gamma}{3\mathrm{\Gamma}}.
\label{eq:locus1}
\end{equation}

\noindent We can further deduce (details in Appendix B)
\begin{equation}
\left.
\begin{IEEEeqnarraybox}[
\IEEEeqnarraystrutmode
\IEEEeqnarraystrutsizeadd{7pt}
{7pt}
][c]{rCl}
u &=& \pm \sqrt{\frac{2 \gamma}{3 \mathrm{\Gamma}} \left[ 1 \pm \tanh(4 \gamma z)  \right]}, \\
v &=& \pm \sqrt{\frac{2 \gamma}{3 \mathrm{\Gamma}} \left[ 1 \mp \tanh(4 \gamma z)   \right]}. 
\end{IEEEeqnarraybox}
\, \right\}
\end{equation}

\noindent These solutions represent four soliton trajectories in all four quadrants of the phase portrait.

How could we explain these soliton solutions on the ground of topology? To this end, we calculate effective stiffness for each spring along the chain for these soliton solutions. We find them remarkably close to each other, as shown in Fig.~\ref{fig3}(c). Though numerically-extracted profile does not represent the exact soliton solution, we claim that this soliton solution indicates that the ``local bandgap'', dictated by two consecutive effective stiffnesses, \textit{closes} along the entire chain. This, therefore, hints at a local topological transition as per Eq.~\eqref{eq:zak}. We can prove it analytically in the following way. For displacements $u$ and $v$ inside a unit cell, its neighboring unit cells would have the displacements $-(u \pm u')$ and $-(v \pm v')$. Strain amplitudes in two consecutive springs would thus be $\mathrm{\Delta}A (\textrm{softening})=|u-v|$ and  $\mathrm{\Delta}A (\textrm{stiffening})=|u+v + u'|$ or $|u+v -v'|$. For the local bandgap to close, we need $\text{K}_{\text{stiffening}}  = \text{K}_{\text{softening}}$ for this unit cell. Therefore, from Eq.~\eqref{eq:es}, we deduce
\begin{equation}
2 \gamma - \frac{3}{4} \mathrm{\Gamma}(u-v)^2 - \frac{3}{4} \mathrm{\Gamma}  (u+v + u')^2 = 0
\end{equation}

\noindent Since $\gamma$, $\mathrm{\Gamma}$, and $u'$ are small quantities, we neglect their higher power and deduce the condition for the closure of local bandgap
\begin{equation}
u^2+v^2= \frac{4 \gamma}{3\mathrm{\Gamma}}.
\end{equation}

\noindent This condition is exactly the same as soliton locus in the phase portrait given by Eq.~\eqref{eq:locus1} and proves the claim that soliton profiles allow effective stiffness distribution in such a way that local bandgap closes along the entire chain. On a more intuitive level, we see the soliton in  Fig.~\ref{fig3}(b) as a solution consisting of two \textit{plane} waves towards the left and the right extremities, and those individually represent the closure of the bandgap. We conjecture that the soliton solution can be thought as a symmetry-breaking solution that transitions between these two plane waves of \textit{opposite} polarities and maintains the bandgap closure, locally.
It is straightforward to conclude that we would have $\text{K}_{\text{stiffening}}  < \text{K}_{\text{softening}}$, locally, along the chain for $u^2+v^2 < {4 \gamma}/{3\mathrm{\Gamma}}$. This is the region inside the circle and corresponds to small amplitudes. The region is topologically trivial with $\mathcal{Z}=0$. In the same way, the region outside the circle, representing large amplitudes, could be shown to have $\text{K}_{\text{stiffening}}  > \text{K}_{\text{softening}}$, locally, along the chain. This would be topologically-nontrivial with $\mathcal{Z}=\pi$.

\subsection{Topologically-trivial solutions}
For this trivial bandgap, we focus on evanescent solutions (in red). These connect the origin $(u,v)=(0,0)$ to neighboring 4 fixed points in Fig.~\ref{fig3}(a). Their profiles would decay to zero amplitude in one direction, similar to the linear evanescent mode shown in 
Appendix C, but saturate to a constant in the opposite direction. This constant is dictated by a fixed point in the phase portrait. It is easy to verify that the strain amplitude corresponding to this fixed point is the same $\D A_0$ we see in Fig.~\ref{fig2}(b). We extract one such solution with the initial condition at $(u_0,v_0)=(0, 0.98)$ and plot displacements in Fig.~\ref{fig3}(d). We can also derive
an analytical form of this solution by substituting $u=0$ in Eq.~\eqref{eq:portrait} and solving for the initial condition $(u,v)=(0, v_0)$ at $z=z_0$ (details in Appendix B), and we get 
\begin{equation}
\left.
\begin{IEEEeqnarraybox}[
\IEEEeqnarraystrutmode
\IEEEeqnarraystrutsizeadd{7pt}
{2pt}
][c]{rCl}
u &=& 0, \\
v &=&\pm  \left[ \frac{3 \mathrm{\Gamma}}{4 \gamma }  + \left (\frac{1}{v_0^2} - \frac{3 \mathrm{\Gamma}}{4 \gamma } \right) \exp[4 \gamma (z-z_0)] \right ]^{-\frac{1}{2}}.
\end{IEEEeqnarraybox}
\, \right\}
\end{equation}
\noindent Clearly, for the initial condition $|v_0| < 2 \sqrt{\gamma / 3\mathrm{\Gamma}}$, we have $|v|$ decreasing to zero as $z\rightarrow \infty$. In the linear limit ($\mathrm{\Gamma} \rightarrow 0$), this mode profile converges to the conventional evanescent solutions observed in linear systems.
We can easily calculate strain amplitude in each spring for this mode and verify $\text{K}_{\text{stiffening}}  < \text{K}_{\text{softening}}$ along the entire chain [Fig.~\ref{fig3}(e)] using Eq.~\eqref{eq:es}, confirming a topologically-trivial solution. See also comment on four other fixed points and neighboring periodic orbits within this region of the phase portrait in Appendix D.

 \subsection{Topologically-nontrivial solutions}
We now discuss the large amplitude solutions lying outside the circle in the phase portrait. For example, the trajectory in blue that lies on $v=0$ axis emerges from $u=\pm \infty$ and asymptotically converges to the fixed points $(u,v)=(\pm 2 \sqrt{\gamma / 3\mathrm{\Gamma}}, 0)$, whereas the trajectory in blue that lies on $u=0$ axis emerges from the fixed points $(u,v)=(0, \pm 2 \sqrt{\gamma / 3\mathrm{\Gamma}})$ asymptotically converges to $v=\pm \infty$. To visualize one such solution, we take $(u_0,v_0)=(3,0)$ as an initial condition and plot displacements in Fig.~\ref{fig3}(f). Different from the evanescent solutions in linear systems that exponentially decay to zero in the bulk (see Appendix C), here we see another unique nonlinear solution that decays to form a plateau in the bulk.  Strain amplitude at this plateau is the same as $\D A_0$  in Fig.~\ref{fig2}(b). Such a mode is observed in a photonic lattice with alternating linear and nonlinear interactions in the SSH chain~\citep{Hadad2016}. We can also derive
an analytical form of this solution by substituting $v=0$ in Eq.~\eqref{eq:portrait} and solving for the initial condition $(u,v)=(u_0,0)$ at $z=z_0$ (details in Appendix B), and it is given by
\begin{equation}
\left.
\begin{IEEEeqnarraybox}[
\IEEEeqnarraystrutmode
\IEEEeqnarraystrutsizeadd{7pt}
{2pt}
][c]{rCl}
u &=& \pm \left[ \frac{3 \mathrm{\Gamma}}{4 \gamma }  + \left (\frac{1}{u_0^2} - \frac{3 \mathrm{\Gamma}}{4 \gamma } \right) \exp[-4 \gamma (z-z_0)] \right ]^{-\frac{1}{2}}, \\
v &=& 0,
\end{IEEEeqnarraybox}
\, \right\}
\end{equation}

\noindent where the negative sign refers to another solution in the phase portrait with negative $u_0$ as the initial condition. We can verify that for the initial condition $|u_0| > 2 \sqrt{\gamma / 3\mathrm{\Gamma}}$,  $|u|$ decreases when $z$ increases, and it saturates to $2 \sqrt{\gamma / 3\mathrm{\Gamma}}$ for $z\rightarrow \infty$. There is no linear limit for such a nonlinear mode since for $\mathrm{\Gamma} \rightarrow 0$ the mode is pushed to infinite amplitude with a plateau at infinite. 
Upon calculating strain amplitude in each spring, we use Eq.~\eqref{eq:es} to plot effective stiffness along the chain in Fig.~\ref{fig3}(g). As discussed above, we verify that $\text{K}_{\text{stiffening}}  > \text{K}_{\text{softening}}$ all along the chain, confirming a topologically-nontrivial solution.

Therefore, we can derive a novel interpretation of topology directly from the phase portrait and show as a schematic in Fig.~\ref{fig3}(h). For small amplitudes, near the origin, the bulk has a topologically-trivial bandgap. It hosts a unique nonlinear evanescent mode shown in Fig.~\ref{fig3}(d).
Whereas for large amplitudes, the bulk has topologically-nontrivial bandgap containing a different looking evanescent mode as shown in Fig.~\ref{fig3}(f). 
Finally, for intermediate amplitudes, a band inversion occurs such that there exist soliton solutions that locally close the bandgap along the chain, thereby forming a ``self-induced'' domain wall between two topologically-distinct regions. Note that the topological classification is dependent on the choice of the unit cell here. By flipping the springs in the unit cell, the region inside (outside) the soliton trajectory in the phase portrait would change to a topologically-nontrivial (trivial) region. The essential point is to recognize the topological \textit{distinction} between the outer and the inner regions. In the following section, we would use full numerical simulations to demonstrate the existence of the bulk nonlinear solutions seen in this section for finite chain settings.

\section{Numerical verification}  \label{Section5}
We enforce a fixed-fixed boundary condition and take 299 masses along the chain. We treat the system semi-infinite, and therefore, we only focus on the edge solution on the left boundary. We first consider a case of a symmetry-preserving boundary, in which the boundary does \textit{not} cut our chosen unit cell.
We, therefore, expect that the bulk-boundary correspondence would apply and only the topologically-nontrivial bulk would support an edge mode \citep{Delplace2011,Prodan2016}.
We use the solutions shown in Figs.~\ref{fig3}(b), \ref{fig3}(d), and \ref{fig3}(f) as initial conditions for the chain and numerically solve Eq.~\eqref{eq:eom}. 

 \begin{figure}[t]
\centering
\includegraphics[width=3.3in]{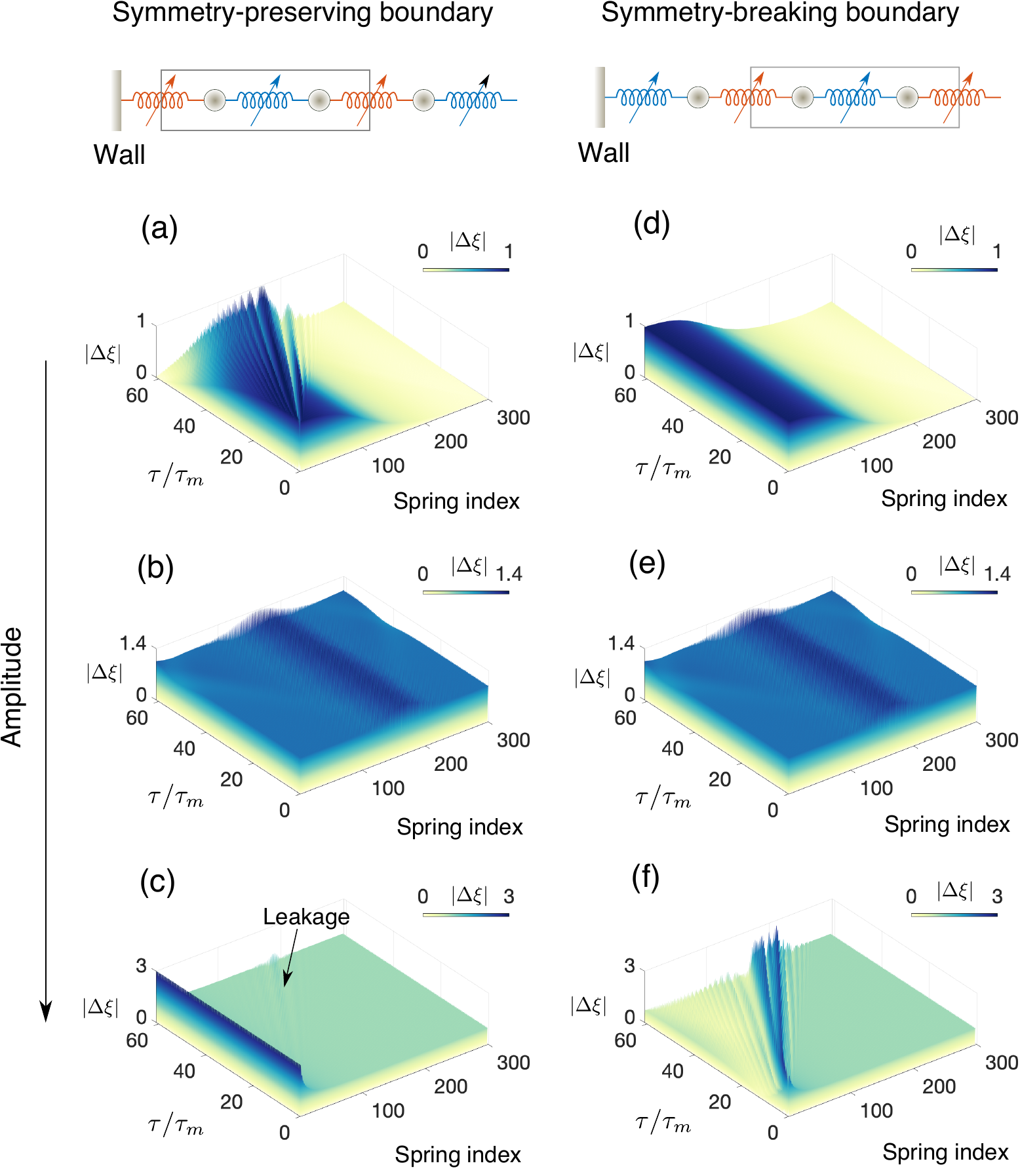}
\caption{[Color online] Full numerical simulations verifying which bulk solutions exist in a finite chain with specific boundary condition. (a), (b), and (c) Solutions for the boundary condition that preserves the lattice symmetry. These show space-time evolution of the absolute value of strain $|\mathrm{\Delta}\xi|$ with the initial conditions obtained from the bulk solutions in Fig.~\ref{fig3}(d), Fig.~\ref{fig3}(b), and Fig.~\ref{fig3}(f), respectively. (d), (e), and (f) The same but for the boundary condition that breaks the lattice symmetry. $\tau_m = 2 \pi/\mathrm{\Omega}_m$ represents the time period of oscillations at the mid-gap frequency. Spring index 1 denotes the spring attached to the wall. Note that we are not interested in seeing the effect of boundary on the soliton solutions in (b) and (e), hence those are kept the same.}
\label{fig4}
\end{figure} 

Figures~\ref{fig4}(a)-(c) show the resulting spatiotemporal diagram for the absolute value of strain in each spring. We choose this quantity for better visualization of 3D plots with minimal oscillatory components. 
The mode profile [Fig. \ref{fig3}(d)], representing a small amplitude regime, shows significant scattering in its time evolution in Fig.~\ref{fig4}(a). As predicted before, this evanescent solution belongs to a topologically-trivial bulk, and therefore, it does not lead to a nonlinear mode for this finite chain. 
We then show the time evolution of the soliton profile [Fig.~\ref{fig3}(b)] in Fig.~\ref{fig4}(b). This represents an intermediate amplitude regime. We do not observe any significant radiation in this case, confirming that this is a solution of the finite chain. We assign the small radiation to the error in extracting the exact soliton trajectory from the phase portrait numerically. 
Finally, for large amplitude regimes, we plot the space-time evolution of the mode profile [Fig.~\ref{fig3}(f)] in Fig.~\ref{fig4}(c). It is reasonably stable initially, and thus, it confirms the existence of this mode for the finite chain. This demonstrates the bulk-boundary correspondence as the bulk is topologically-\textit{nontrivial} in this regime. We do observe some leakage emanating from the sharp localized peak on the boundary. We assign this error to the fact that the mode profile is obtained using continuum approximation, which assumes a slow variation of displacement amplitude along the chain. Therefore, it is expected to be less accurate for the sharp localized peak on the boundary. 

Now we consider a finite chain with a symmetry-breaking boundary condition, which cuts our chosen unit cell in half. This case is equivalent to having a symmetry-preserving boundary condition, but with a ``new'' unit cell that has two types of springs flipped. Therefore, we expect only the topologically-\textit{trivial} solution shown in Fig. \ref{fig3}(d) would lead to a nonlinear edge solution.
 In Figs. \ref{fig4}(d)-(f), we plot the time evolution of the three mode profiles for such a boundary. Clearly, we observe just the opposite trend. 
We see that the evanescent mode, belonging to a small amplitude regime, does lead to a nonlinear mode of the finite chain as shown in Fig.~\ref{fig4}(d). However, the evanescent mode, belonging to a large amplitude regime, shows significant scattering, and thus, it does not lead to a nonlinear mode as shown in Fig.~\ref{fig4}(f). The evolution of the soliton profile in Fig.~\ref{fig4}(e) should remain the same as it is away from the boundaries and we do not change the boundary conditions in that case. 

In this way, we have demonstrated that one can utilize amplitude dependency in the system to realize nonlinear edge solutions for finite chains. 
Taking this idea forward, could we also demonstrate such a transition \textit{in situ}, for possible experimental settings, where only boundaries are usually excited? 
Focusing only on symmetry-preserving boundary conditions hereafter, could we transition to a topologically-nontrivial solution and excite the nonlinear edge mode of large amplitudes?
For that to happen, a significant amount of transient excitation would be needed so that one can induce the given mode profile. This is where the system with a cubic type of nonlinear profile [cf. Eq.~\eqref{eq:cubic}] might see some limitations because the high instantaneous strain in spring could make its effective stiffness close to zero, and thus lead to stability issues. Therefore, building on the findings so far, in the latter part of this paper, we propose another nonlinear force-deformation law for springs that circumvents this issue and demonstrate a ``self-induced'' edge mode by using a boundary excitation.

\section{Saturable nonlinearity: A good alternative for experiments}  \label{Section6}
\begin{figure}[t]
\centering
\includegraphics[width=3.3in]{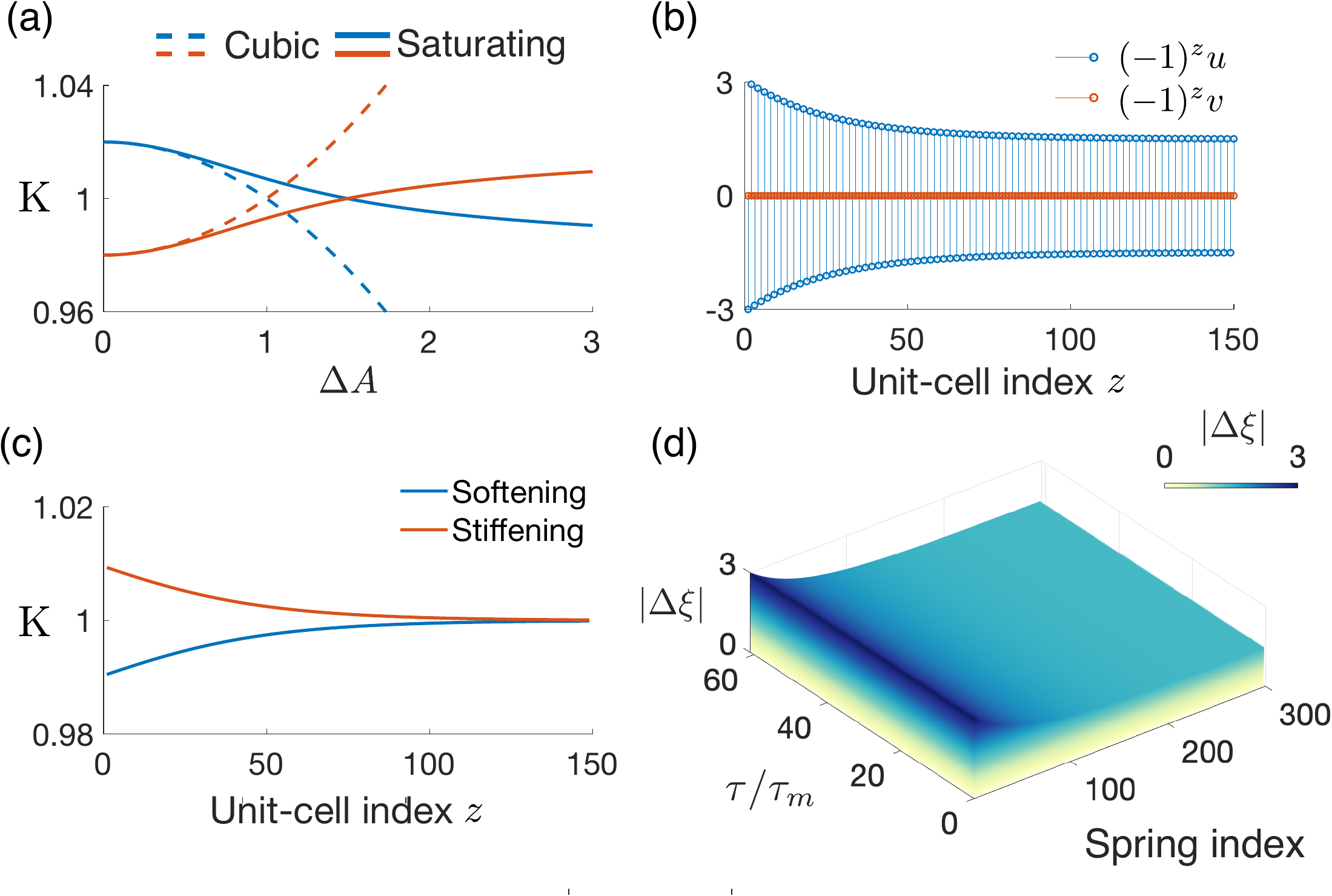}
\caption{[Color online] (a) Effective stiffness of nonlinear springs as a function of strain amplitude $\mathrm{\Delta}A$ when we change nonlinearity type from cubic to saturating. 
(b) Resulting profile of the edge mode. 
(c) Effective stiffness of springs along the chain for this mode.
(d) Full numerical simulation showing space-time evolution of the absolute value of strain $|\mathrm{\Delta}\xi|$ when we take the edge mode profile as the initial condition in a finite chain with the symmetry-preserving boundary.}
\label{fig5}
\end{figure}  

\subsection{Effective stiffness}
We take the same spring-mass system shown in Fig.~\ref{fig1}(b), but replace spring force-deformation law as 
\begin{equation}
\left.
\begin{IEEEeqnarraybox}[
\IEEEeqnarraystrutmode
\IEEEeqnarraystrutsizeadd{7pt}
{7pt}
][c]{rCl}
\text{F}( \mathrm{\Delta} x)&=(1+\gamma) k  \mathrm{\Delta} x - 2 \gamma k \frac{\erf(\nu  \mathrm{\Delta} x)}{c \nu}    \; \;  \textrm{(stiffening)}, \\
\text{F}( \mathrm{\Delta} x)&=(1-\gamma) k  \mathrm{\Delta} x+ 2 \gamma k \frac{\erf(\nu  \mathrm{\Delta} x)}{c \nu}    \; \;  \textrm{(softening)},
\end{IEEEeqnarraybox}
\, \right\}
\label{eq:saturating}
\end{equation}

\noindent where ``erf" denotes the error function, $\nu$ is a parameter to tune its profile, and $c=2/\sqrt{\pi}$. One can verify that stiffening spring follows $\text{F}( \mathrm{\Delta} x) \approx (1-\gamma)k  \mathrm{\Delta} x$ for small $ \mathrm{\Delta} x$ since $\erf(\nu  \mathrm{\Delta} x) \approx c \nu  \mathrm{\Delta} x$, whereas for large $ \mathrm{\Delta} x$, $\text{F}( \mathrm{\Delta} x) \approx (1+\gamma) k  \mathrm{\Delta} x -  2 \gamma k / c \nu $. This indicates an increase in quasi-static stiffness, i.e., stiffening, which eventually saturates to a constant. We can make a similar argument for softening type of nonlinearity as well.

We now expand the error function as a power series and perform non-dimensionalization on the equations of motion. We thus obtain the following effective stiffness for nonlinear springs as a function of strain amplitude $\mathrm{\Delta}A$ (details in Appendix E)
\begin{equation}
\left.
\begin{IEEEeqnarraybox}[
\IEEEeqnarraystrutmode
\IEEEeqnarraystrutsizeadd{7pt}
{7pt}
][c]{rCl}
\text{K}_{\textrm{stiffening}}(\mathrm{\Delta}A) &=&(1-\gamma)    \\ 
&&+\> \sum_{n=3,5,..} (-1)^{\frac{n+1}{2}} n C_{\frac{n-1}{2}}  \mathrm{\Gamma}_n   \bigg (\frac{\mathrm{\Delta}A}{2} \bigg )^{n-1}, \\ 
\text{K}_{\textrm{softening}}(\mathrm{\Delta}A) &=&(1+\gamma)   \\ 
&&-\> \sum_{n=3,5,..} (-1)^{\frac{n+1}{2}} n C_{\frac{n-1}{2}}  \mathrm{\Gamma}_n   \bigg (\frac{\mathrm{\Delta}A}{2} \bigg )^{n-1} , 
\end{IEEEeqnarraybox}
\, \right\},
\label{eq:es_sat}
\end{equation}

\noindent where $C$ represents the Catalan number and $\mathrm{\Gamma}_n$ is the nonlinearity parameter for the $n$th-order nonlinearity. For a better comparison, we choose $\nu$ in the force profile such that the first nonlinearity parameter $\mathrm{\Gamma}_3$ equals $\mathrm{\Gamma}$ of the cubic case in the first half of the paper.

In Fig.~\ref{fig5}(a), we plot effective stiffness calculated from Eq.~\eqref{eq:es_sat} and compare it with the cubic case in Eq.~\eqref{eq:es}. We take first 60 terms in the infinite series for the numerical calculation. For small $\mathrm{\Delta}A$, both types of profiles are close to each other since $\mathrm{\Gamma}_3=\mathrm{\Gamma}$. But, as the amplitude increases, higher order nonlinear terms in saturating profile become dominant, reflecting as the saturation in effective stiffness.  We also observe that the saturating profiles also cross each other at a critical amplitude  $\mathrm{\Delta}A_0 \approx 1.5$, which is \textit{larger} than the one for cubic profiles ($\mathrm{\Delta}A_0 = 1$). This, therefore, would reflect as a different plateau level for the edge mode.

\begin{figure*}[t]
\centering
\includegraphics[width=6in]{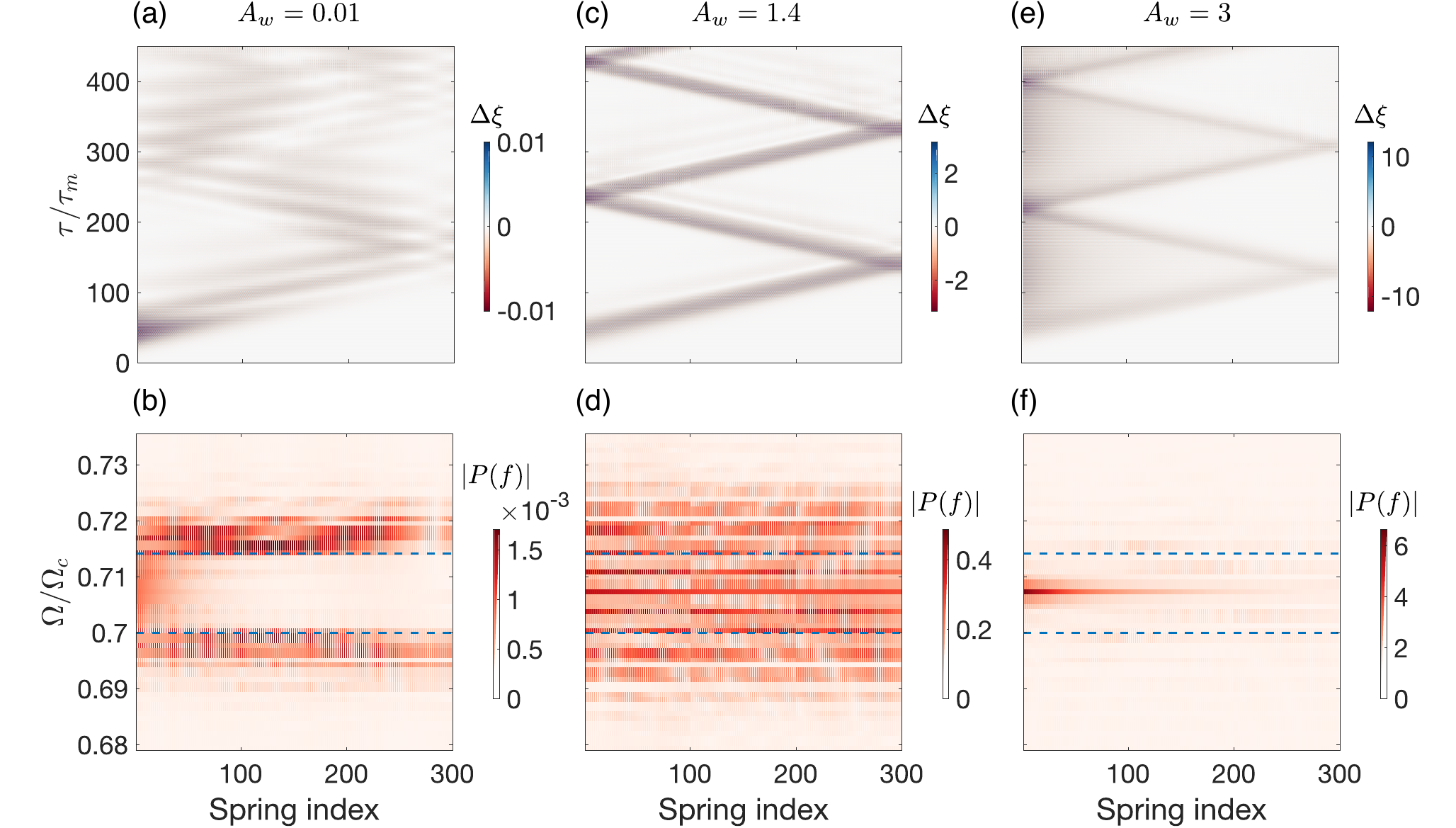}
\caption{[Color online] (a) Space-time evolution of strain $\mathrm{\Delta}\xi$ when we excite the left wall with a Gaussian-modulated profile of  displacement amplitude $A_w=0.01$. 
(b) Spectrum calculated from the time history of spring strains for the case above.
(c)-(d) The same with displacement amplitude $A_w=1.4$.
(e)-(f) The same with displacement amplitude $A_w=3$.  Spectrum plots in (b), (d), and (f) indicate the presence of a bandgap, its closure, and the emergence of an edge mode, respectively. Blue dashed lines mark the lower and upper band edges for the initial (linear) configuration. $\mathrm{\Omega}_c$ denotes the upper cutoff frequency of the optical branch.}
\label{fig6}
\end{figure*} 

\subsection{Edge solutions}
To get the evanescent mode profile in this case, we would again use continuum approximation. Due to its complexity, we do not calculate the entire phase portrait. Rather, we extract the dynamical equations only for $u=0$ or $v=0$, on which the evanescent modes lie [see Fig.~\ref{fig3}(a)]. For example, for $v=0$ we obtain
\begin{equation}
\left.
\begin{IEEEeqnarraybox}[
\IEEEeqnarraystrutmode
\IEEEeqnarraystrutsizeadd{2pt}
{2pt}
][c]{rCl}
u'&=&2 \gamma u + 2 u  \big ( - {\mathrm{\Lambda}_3}u^2 + {\mathrm{\Lambda}_5}u^4 - {\mathrm{\Lambda}_7}u^6 + ... \big)  \\
v'&=&0 
\end{IEEEeqnarraybox}
\, \right\},
\label{eq:edge_sat}
\end{equation}

\noindent where $\mathrm{\Lambda}_{n} =  n C_{\frac{n-1}{2}} \mathrm{\Gamma}_n/2^{n-1}$ with $n=\{3,5,7,...\}$. It is easy to verify that when higher order nonlinearity is ignored ($\mathrm{\Gamma}_{n}=0$ for $n \geq 5$), the above equation reduces to Eq.~\eqref{eq:portrait} with $\mathrm{\Gamma}_3=\mathrm{\Gamma}$ and $v=0$. 

In Fig.~\ref{fig5}(b), we plot the evanescent mode profile obtained from these equations with the initial condition $(u_0,v_0)=(3,0)$. We clearly see a decaying mode from the left that eventually saturates to a plateau, similar to the mode profile earlier in Fig.~\ref{fig3}(f). Again, this is the implication of the nontrivial topology of the bulk, in which the plateau level is explained by the existence of a fixed point, which corresponds to strain amplitude $\mathrm{\Delta}A_0 \approx 1.5$ in this case. In Fig.~\ref{fig5}(c), we plot the effective stiffness in each spring for the aforementioned mode profile, and it is clear that $\text{K}_{\text{stiffening}} > \text{K}_{\text{softening}}$, locally, for the entire chain. By performing full numerical simulations and plotting in Fig.~\ref{fig5}(c), we also confirm that this evanescent mode is a solution of a finite chain with the symmetry-preserving boundary condition.

\subsection{In situ emergence of an edge mode}  
In this section, we numerically demonstrate how an in situ topological phase transition could be achieved by increasing the excitation amplitude given at one side of the chain. As a result, we will show the emergence of an edge mode for high amplitude excitation. We apply a Gaussian-modulated sine signal in displacement, centered at the mid-gap frequency, to the left boundary wall of the chain. This is given as $u_w(\tau) =A_w \sin(\mathrm{\Omega}_m \tau) \exp \big[-\frac{1}{2} (\frac{\tau-200}{70})^2 \big]$. In Fig.~\ref{fig6}, we plot space-time evolution of strain and its frequency spectrum for three different $A_w$ values. For small excitation amplitude $A_w=0.01$, we see in Fig.~\ref{fig6}(a) that the propagating wave is dispersive. By looking at the corresponding frequency spectrum in Fig.~\ref{fig6}(b) calculated from time-history of strain, we observe the existence of a bandgap. The width of bandgap also matches with the blue dotted lines, which are analytically-obtained cutoff frequencies in a linear dimer chain with spring coefficients $1-\gamma$ and $1+\gamma$.  We do not observe any energy localization on the left end for the frequencies inside the bandgap, except for the usual evanescent components, indicating topologically \textit{trivial} regime. When we increase amplitude $A_w$ to 1.4, we see a distinctive wave propagation in Fig.~\ref{fig6}(c). The wave propagation is comparatively less dispersive and is in the form of a bundle. This could correspond to a \textit{traveling} soliton solution, in the form of a moving domain wall, which can be analyzed from Eq.~\eqref{eq:dirac}. We will leave this topic for future investigations as we focus only on the stationary solutions in this study. We can see that its spectrum in Fig.~\ref{fig6}(d) does not show any clear bandgap, indicating a transition regime for band topology. Finally, for high amplitude excitation $A_w=3$, we remarkably observe dominant localized vibrations in Fig.~\ref{fig6}(e) on the left side of the chain. This manifests as a clear and sharp frequency peak in the spectrum plot in Fig.~\ref{fig6}(f).

\begin{figure}[t]
\centering
\includegraphics[width=3.3in]{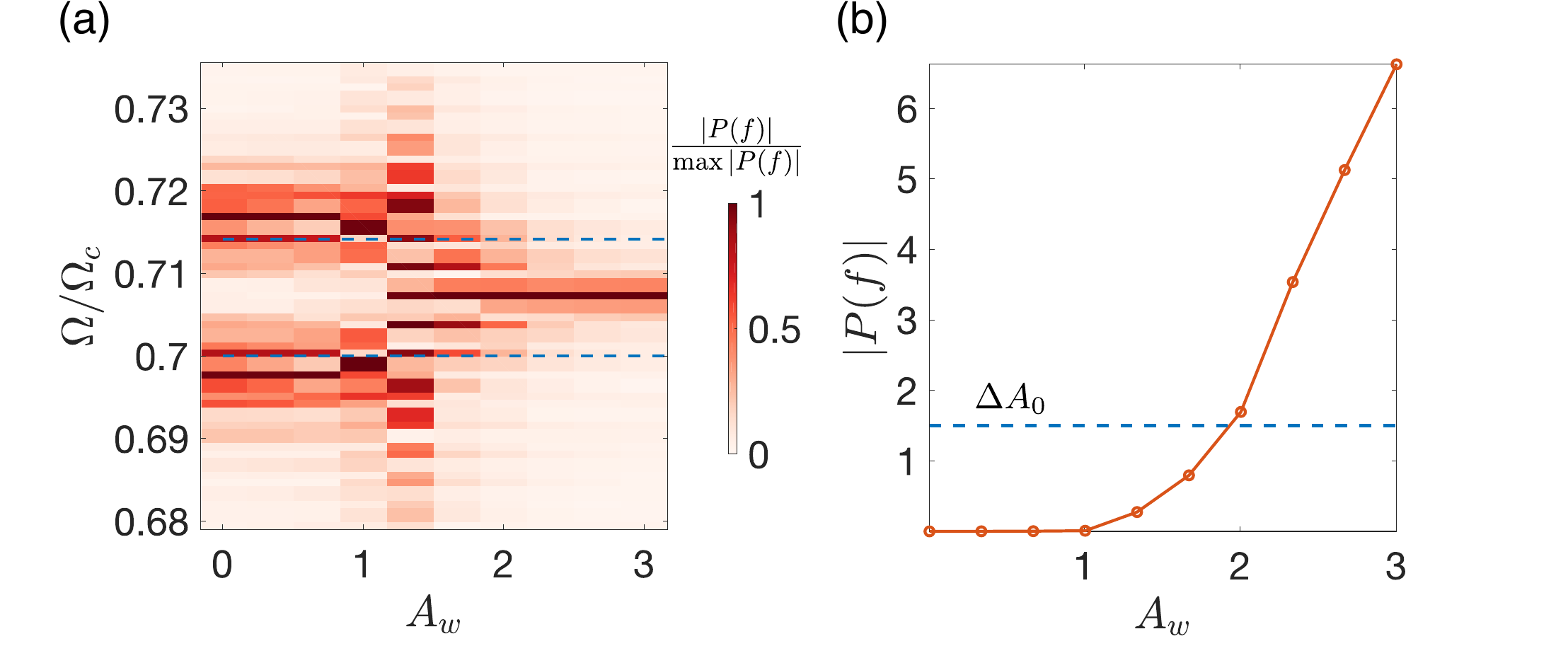}
\caption{[Color online] (a) Normalized modal amplitude of the first spring attached to the wall as a function of excitation amplitude  $A_w$. This shows that the emergence of an edge mode is a result of bandgap closure. (b)  Modal amplitude of the first spring as a function of excitation amplitude at the \textit{mid-gap} frequency. The critical strain amplitude $\mathrm{\Delta}A_0$ is marked to indicate the point of band closure and steep increase in modal amplitude thereafter. }
\label{fig7}
\end{figure}

To see this transition in the spectrum even more clearly, we plot strain spectrum only for the first spring  against a range of wall displacement amplitude in Fig.~\ref{fig7}(a). Increasing $A_w$ also increases the modal amplitude $|P(f)|$, therefore for better visualization, we choose to plot \textit{normalized} modal amplitude w.r.t. the peak amplitude at any frequency for a given wall displacement. In this way, we can reasonably predict and observe the presence of a bandgap, its closure, and the emergence of mid-gap-frequency spectral peak, upon increasing the excitation amplitude. This, therefore, demonstrates excitation-dependent in situ topological transition in the system. In Fig.~\ref{fig7}(b), we plot the modal amplitude at the first spring, but only at the mid-gap frequency. Though a rise in modal amplitude starts around $A_w \approx 1$, we notice that it becomes steeper only after $A_w \approx 2$. This regime is remarkably close to the topological transition strain amplitude $|P(f)|= \mathrm{\Delta}A_0 \approx 1.5$ predicted earlier, which again confirms the fundamental topological mechanism at work. 

\section{Conclusions and future directions}  \label{Section7}
To summarize the main findings of this study, we successfully provide a scheme to engineer a nonlinear phononic crystal that can be transitioned to topologically distinct regimes simply by changing its dynamic excitation. The higher excitation invokes nonlinear effects in the system, and we utilize this effect to close the bandgap of an initially linear system and open it again with amplitude, leading to a band inversion. We rely on a framework based on \textit{effective} stiffness to show that the system makes a topological transition. Using this framework, if topological characterization  is done locally, it provides a powerful tool to explain the emerging nonlinear solutions.

In the first part of this study, we consider \textit{cubic} nonlinearity in alternating ``stiffening" and ``softening" types of springs and track steady-state solutions at the mid-gap frequency using a phase portrait.  We witness circular heteroclinic trajectories that correspond to kink solitons. 
Interestingly, for these solutions ``local bandgap'' closes in the system and thus these represent topological transition points. In the phase portrait, this orbit separates two topologically distinct regimes in small and large amplitude regimes and can be treated as a ``self-induced'' domain wall. We can, therefore, utilize this amplitude dependency to transition from one topological region to another and vice-versa. Depending on the choice of the unit cell and boundary, one of these regimes would be topologically nontrivial and would host a unique nonlinear edge mode, not witnessed in linear systems. Note that these observations are different from the ones reported recently in circuits~\citep{Hadad2018a}, where \textit{a line of fixed points} separated two topologically distinct regimes. This difference, in addition to the existence of more fixed points, in our case is due to the fact that we have nonlinearity in all the springs, and that gives an extra degree of freedom for the effective stiffness along the chain to vary with amplitude.

In the latter part, we build on the previous findings to suggest a system with \textit{saturating} type of nonlinearity that could potentially be a better candidate for experiments. We show that simply by exciting one end of the chain, the system transitions to a topologically-nontrivial regime and thus emerges a nonlinear edge mode for large amplitudes. In principle, this nonlinearity management could be realized by utilizing geometric nonlinearity in tunable systems or performing a rigorous topology optimization of a mechanical structure. 

It would be interesting to examine the stability of such nonlinear solutions in future studies. Especially, the high-amplitude evanescent modes and kink solitons might radiate to low-amplitude evanescent modes in the bulk, leading to semistable behavior with a finite basin of attraction in the phase portrait~\cite{Leykam2016}. If there exist certain stability in modes, how does it relates to the robustness, which is often associated with topological systems? How does discreteness affect these solutions? Could we have \textit{traveling} soliton solutions in the system as Fig.~\ref{fig6}c seems to suggest? It would also be interesting to extend this framework to higher dimensions, where we might expect a completely new family of solutions not witnessed in linear systems.

\begin{acknowledgments}
We thank Professor Dimitri Frantzeskakis for fruitful discussions. This work has been supported by the project CS.MICRO funded under the program Etoiles Montantes of the Region Pays de la Loire. 
\end{acknowledgments}


\appendix

\renewcommand{\appendixname}{APPENDIX}
\setlength\parskip{0.5\baselineskip}

\section{BAND-EDGE FREQUENCIES AND EFFECTIVE STIFFNESS}
We get  the nonlinear continuation of lower band-edge frequency by substituting $u=v=A$ ($A$ being the amplitude) in Eq.~\eqref{eq:ode}
 \begin{eqnarray}
\mathrm{\Omega}_l^2=2 (1-\gamma)+6   \mathrm{\Gamma} A^2.  \nonumber
\end{eqnarray}

\noindent Clearly, this band-edge frequency increases with amplitude. This makes sense because for this mode, softening type nonlinear spring does not see any strain, rather only stiffening type spring sees a strain of amplitude $\mathrm{\Delta} A = 2 A$ [verify this by looking at the unit cell in Fig.~\ref{fig1}(b)]. Frequency of the oscillations can thus be written in terms of \textit{effective} stiffness and \textit{effective} mass as $\w_l^2 = \text{K}_{\textrm{stiffening}}/m_{\textrm{eff}}$, where  $m_{\textrm{eff}}=1/2$ for unit mass chain. We can thus derive {effective} stiffness for this stiffening spring in terms of strain as
\begin{eqnarray}
\text{K}_{\textrm{stiffening}}(\mathrm{\Delta}A) = \frac{\mathrm{\Omega}_l^2}{2} =1-\gamma +\frac{3}{4}  \mathrm{\Gamma} (\mathrm{\Delta}A)^2 . 
\label{eq:esStiff}
\end{eqnarray}

Similarly, we substitute $u=-v=A$ in Eq.~\eqref{eq:ode} to get the upper band-edge frequency 
\begin{eqnarray}
\mathrm{\Omega}_u^2=2 (1+\gamma) -6 \mathrm{\Gamma} A^2. \nonumber
\end{eqnarray}

\noindent We see that this band-edge frequency decreases with amplitude. This also makes sense because for this mode, stiffening type nonlinear spring does not see any strain, rather only softening type spring sees a strain of amplitude $\mathrm{\Delta}A=2 A$ [this can also be verified by looking at the unit cell in Fig.~\ref{fig1}(b)]. We can thus define effective stiffness for this softening spring in terms of strain as 
\begin{eqnarray}
\text{K}_{\textrm{softening}}(\mathrm{\Delta}A) = \frac{\mathrm{\Omega}_u^2}{2} = 1+\gamma -\frac{3}{4} \mathrm{\Gamma} (\mathrm{\Delta}A)^2. 
\label{eq:esSoft}
\end{eqnarray}

Now one might ask if this effective stiffness framework is valid for \textit{all} the solutions inside the bandgap? The answer is Yes, especially here when we have a small bandgap and we are still looking for Bloch solutions. One of the ways to check this is to consider a \textit{linear} dimer system with two springs with stiffness $\text{K}_{\textrm{softening}}$ and $\text{K}_{\textrm{stiffneing}}$ for the unit cell shown in Fig.~\ref{fig1}(b). Let $\text{K}_{\textrm{softening}} > \text{K}_{\textrm{softening}}$. We then follow the same approach discussed in Section~\ref{Section3} to derive dynamical equations for excitation frequency $\w$ as

\begin{equation}
\left.
\begin{IEEEeqnarraybox}[
\IEEEeqnarraystrutmode
\IEEEeqnarraystrutsizeadd{7pt}
{7pt}
][c]{rCl}
\text{K}_{\textrm{stiffening}} u'&=& - (\text{K}_{\textrm{softening}} + \text{K}_{\textrm{stiffening}} -\w^2)v \\
&&+\> (\text{K}_{\textrm{softening}}-\text{K}_{\textrm{stiffening}}) u, \\
 \text{K}_{\textrm{stiffening}} v'&=&  (\text{K}_{\textrm{softening}} + \text{K}_{\textrm{stiffening}}-\w^2)u \\
&&-\> (\text{K}_{\textrm{softening}}-\text{K}_{\textrm{stiffening}} )v. 
\end{IEEEeqnarraybox}
\, \right\}
\label{eq:portraitES}
\end{equation}

\noindent We then substitute effective stiffness expression given by Eq.~\eqref{eq:esStiff} and  Eq.~\eqref{eq:esSoft} in  Eq.~\eqref{eq:portraitES} to obtain
\begin{equation}
\left.
\begin{IEEEeqnarraybox}[
\IEEEeqnarraystrutmode
\IEEEeqnarraystrutsizeadd{7pt}
{7pt}
][c]{rCl}
\IEEEeqnarraymulticol{3}{l}{
\left[ 1- \gamma + \frac{3}{4}  \mathrm{\Gamma}  (\mathrm{\Delta}A_{\textrm{stiffening}})^2 \right] u' 
}\nonumber\\
&=& - \left [2 +\frac{3}{4}  \mathrm{\Gamma} \left (\mathrm{\Delta}A_{\textrm{stiffening}} - \mathrm{\Delta}A_{\textrm{softening}} \right )^2  -\w^2 \right] v \\
&&+\>  \left [2 \gamma - \frac{3}{4}  \mathrm{\Gamma} \left (\mathrm{\Delta}A_{\textrm{stiffening}} + \mathrm{\Delta}A_{\textrm{softening}} \right )^2   \right] u, \\
\IEEEeqnarraymulticol{3}{l}{
\left[ 1- \gamma + \frac{3}{4}  \mathrm{\Gamma}  (\mathrm{\Delta}A_{\textrm{stiffening}})^2 \right] v' 
}\nonumber\\
&=&  \left [2 +\frac{3}{4}  \mathrm{\Gamma} \left (\mathrm{\Delta}A_{\textrm{stiffening}} - \mathrm{\Delta}A_{\textrm{softening}} \right )^2  -\w^2 \right] u \\
&&-\>  \left [2 \gamma - \frac{3}{4}  \mathrm{\Gamma} \left (\mathrm{\Delta}A_{\textrm{stiffening}} + \mathrm{\Delta}A_{\textrm{softening}} \right )^2   \right] v.
\end{IEEEeqnarraybox}
\, \right\}
\end{equation}

\noindent Using continuum approximation, we can estimate the general strain amplitude in two consecutive springs as $\mathrm{\Delta}A_{\textrm{softening}}=|u-v|$ and  $\mathrm{\Delta}A_{\textrm{stiffening}}=|u+v + u'|$ or $|u +v - v'|$. These we substitute in the equations above and neglect  higher order terms for small $\gamma$, $\mathrm{\Gamma}$, $u'$, and $v'$ 
\begin{equation}
\left.
\begin{IEEEeqnarraybox}[
\IEEEeqnarraystrutmode
\IEEEeqnarraystrutsizeadd{7pt}
{7pt}
][c]{rCl}
u'&=& - (2-\w^2)v + 2 \gamma u - \frac{3}{2}\mathrm{\Gamma}(u^3+3u v^2), \\
v'&=&  (2-\w^2)u - 2 \gamma v + \frac{3}{2}\mathrm{\Gamma}(v^3+3u^2 v),
\end{IEEEeqnarraybox}
\, \right\}
\end{equation}
 
 \noindent which are exactly the same as Eq.~\eqref{eq:ode} and proves that a linear dynamical system with effective stiffness could be used as a mapping for the nonlinear dynamical system for \textit{all} the frequencies inside the bandgap.
 
\section{ANALYTICAL EXPRESSIONS FOR SOLITON AND EVANESCENT SOLUTIONS}
\textit{Soliton solutions:} An integral of motion of the dynamical system given by Eq.~\eqref{eq:portrait} is
\begin{equation}
E =2 \gamma u v \left [1- \frac{3\mathrm{\Gamma}}{4 \gamma}(u^2+v^2) \right].
\label{eq:integral}
\end{equation}

\noindent If we define an auxiliary function $g=u/v$, we can write the following equation
\begin{equation}
\left (\frac{\mathrm{d}g}{\mathrm{d}z} \right)^2 =16  \gamma^2 g^2- 24  \mathrm{\Gamma} g (1+g^2) E.
\label{eq:auxiliary}
\end{equation}

\noindent We can then extract the integral of motion for the solition solution from the phase portrait in Fig.~\ref{fig3}(a). We substitute $(u,v)=(2 \sqrt{\gamma / 3\mathrm{\Gamma}}, 0)$ in Eq.~\eqref{eq:integral} since this point lies in the soliton trajectory. We obtain $E=0$, which we use in Eq.~\eqref{eq:auxiliary} to get
\begin{equation}
\frac{\mathrm{d}g}{\mathrm{d}z}  = \pm 4 \gamma g.
\end{equation}

\noindent We now use the auxiliary function $g(z) = \exp(\pm 4 \gamma z )$ and Eq.~\eqref{eq:integral} to obtain soliton profiles
\begin{equation}
\left.
\begin{IEEEeqnarraybox}[
\IEEEeqnarraystrutmode
\IEEEeqnarraystrutsizeadd{7pt}
{7pt}
][c]{rCl}
u &=& \pm \sqrt{\frac{2 \gamma}{3 \mathrm{\Gamma}} \left[ 1 \pm \tanh(4 \gamma z)  \right]}, \\
v &=& \pm \sqrt{\frac{2 \gamma}{3 \mathrm{\Gamma}} \left[ 1 \mp \tanh(4 \gamma z)   \right]}. 
\end{IEEEeqnarraybox}
\, \right\}
\end{equation}

\textit{Evanescent solutions:} 
Modes shown in Figs.~\ref{fig3}(d),(f) lie on $u=0$ and $v=0$ axes of the phase portrait. Therefore, to find the analytical form the solutions lying on $v=0$ axis, we substitute $v=0$ in Eq.~\eqref{eq:portrait}
\begin{equation}
\left.
\begin{IEEEeqnarraybox}[
\IEEEeqnarraystrutmode
\IEEEeqnarraystrutsizeadd{7pt}
{2pt}
][c]{rCl}
u'&=&2 \gamma u - \frac{3}{2}\mathrm{\Gamma} u^3, \\
v'&=&0.
\end{IEEEeqnarraybox}
\, \right\}
\label{eq:edge_ana}
\end{equation}

\noindent We define a new variable $y=1/u^2$ and transform these equations to
\begin{equation}
\left.
\begin{IEEEeqnarraybox}[
\IEEEeqnarraystrutmode
\IEEEeqnarraystrutsizeadd{7pt}
{2pt}
][c]{rCl}
y'&=& - 4 \gamma y + 3\mathrm{\Gamma}, \\
v'&=&0.
\end{IEEEeqnarraybox}
\, \right\}
\end{equation}

\noindent Clearly, these are two linear differential equations and can be solved easily. With the initial condition $(u,v)=(u_0,0)$ at $z=z_0$, we thus obtain the evanescent solution profile
\begin{equation}
\left.
\begin{IEEEeqnarraybox}[
\IEEEeqnarraystrutmode
\IEEEeqnarraystrutsizeadd{7pt}
{2pt}
][c]{rCl}
u &=& \pm \frac{1}{\sqrt{y}} = \pm \left[ \frac{3 \mathrm{\Gamma}}{4 \gamma }  + \left (\frac{1}{u_0^2} - \frac{3 \mathrm{\Gamma}}{4 \gamma } \right) \exp[-4 \gamma (z-z_0)] \right ]^{-\frac{1}{2}},\\
v &=& 0.
\end{IEEEeqnarraybox}
\, \right\}
\end{equation}

\noindent Note that these represent two solutions that correspond to a positive and negative sign of the initial condition $u_0$. Similarly, we can obtain solutions lying on $u=0$ axis with the initial condition $(u,v)=(0, v_0)$ at $z=z_0$
\begin{equation}
\left.
\begin{IEEEeqnarraybox}[
\IEEEeqnarraystrutmode
\IEEEeqnarraystrutsizeadd{7pt}
{2pt}
][c]{rCl}
u &=& 0, \\
v &=&\pm  \left[ \frac{3 \mathrm{\Gamma}}{4 \gamma }  + \left (\frac{1}{v_0^2} - \frac{3 \mathrm{\Gamma}}{4 \gamma } \right) \exp[4 \gamma (z-z_0)] \right ]^{-\frac{1}{2}}.
\end{IEEEeqnarraybox}
\, \right\}
\end{equation}

\section{PHASE PORTRAIT OF A LINEAR SYSTEM}
Here we discuss the features of the phase portrait obtained for a chain shown in Fig.~\ref{fig1}, but with only linear springs. To this end, we substitute $\mathrm{\Gamma}=0$ in Eq.~\eqref{eq:portrait} and obtain 
\begin{equation}
\left.
\begin{IEEEeqnarraybox}[
\IEEEeqnarraystrutmode
\IEEEeqnarraystrutsizeadd{2pt}
{2pt}
][c]{rCl}
u'&=&2 \gamma u,  \\
v'&=&- 2 \gamma v.  
\end{IEEEeqnarraybox}
\, \right\}
\label{eq:portrait_lin}
\end{equation}

\begin{figure}[t]
\centering
\includegraphics[width=3.3in]{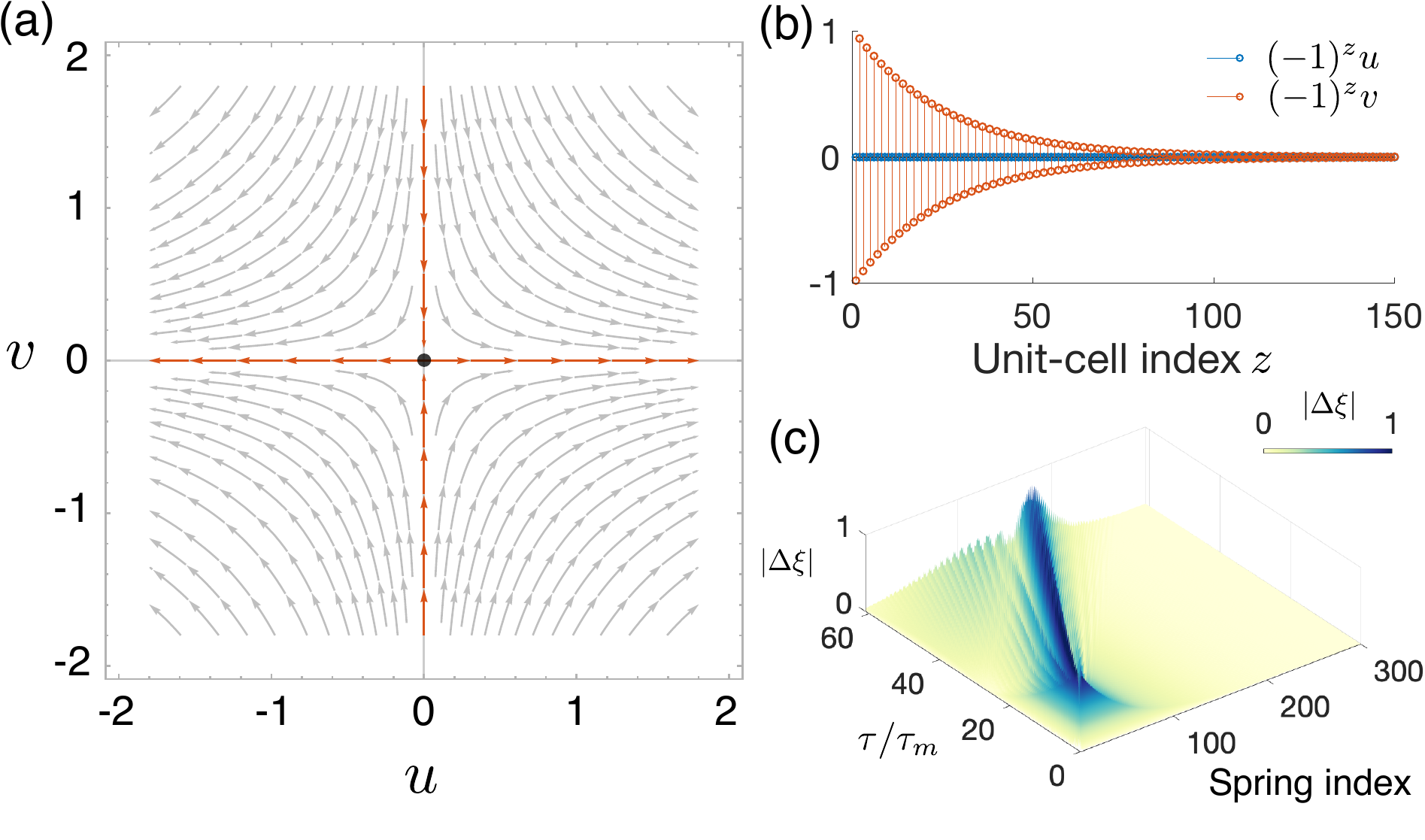}
\caption{[Color online] (a) Phase portrait for a linear system ($\mathrm{\Gamma}=0$) showing steady-state solutions at the mid-gap frequency. There is a fixed point indicated by the black dot at the origin. Red arrows show the evanescent solutions in the bulk. (b) An evanescent mode solution decaying from the left corresponds to the solutions on $u=0$ axis. (c) Space-time evolution of the absolute value of strain when we use the aforementioned mode profile as an initial condition for a finite chain with the symmetry-preserving left boundary.}
\label{fig8}
\end{figure}

\noindent These are two uncoupled first-order differential equations with a fixed point at $(u,v)=(0,0)$. We show the phase portrait in Fig.~\ref{fig8}(a). By comparing it with the one for the nonlinear case in Fig.~\ref{fig3}(a), we notice that its surrounding 8 fixed points tend to go to infinity in the linear limit. As a result, we only see the evanescent modes in red from/to the fixed point at the origin. Thus, we would not expect any ``plateau'' in the mode shape since there is no fixed point at finite nonzero $u$ or $v$. We verify it by extracting a solution with the initial condition $(u_0,v_0)=(0,1)$ and plotting in Fig.~\ref{fig8}(b). We observe that the evanescent solution asymptotically decays to zero displacements. Note that there are other evanescent solutions localized towards the right end, also with the nonlinear case in Fig.~\ref{fig3}(a). We ignore them as our chain is assumed to be semi-infinite and we only focus on the left end as stated in the main text.

This bulk is topologically trivial as one can interpret from Figs.~\ref{fig2}(b), \ref{fig3}(a) in low amplitude (linear) limit. Therefore, the evanescent mode we obtained here is similar to the one in Fig.~\ref{fig3}(d) topologically and this would \textit{not} lead to a nonlinear edge mode for a finite chain with the symmetry-preserving boundary condition. We demonstrate this by full numerical simulation in Fig.~\ref{fig8}(c), which shows significant scattering. 

\section{OTHER FIXED POINTS IN THE PHASE PORTRAIT}
In Fig.~\ref{fig3}(a), we observe 4 fixed points shown in red dots in the region inside the soliton trajectories.  These are $(u,v)=(\pm \sqrt{\gamma/ 3\mathrm{\Gamma}}, \pm \sqrt{\gamma/ 3 \mathrm{\Gamma}})$, which represent periodic solutions for the case when one spring (stiffening or softening) sees zero strain, whereas the other $\pm 2\sqrt{\gamma/ 3\mathrm{\Gamma}}$. As we have two types of nonlinear springs and their strain amplitudes can independently change, so it makes sense that this periodic solution would also lead to the same oscillation frequency ($\mathrm{\Omega}_m$), just as for the other fixed points on $u=0$ and $v=0$ axes.

We also verify that these fixed points, and also their neighboring periodic orbits, would lead to solutions that obey $\text{K}_{\text{stiffening}}  < \text{K}_{\text{softening}}$ for the entire length of the chain. Therefore, this represents a topologically trivial regime, surrounding by soliton trajectories, as also observed for the evanescent solutions in Section~\ref{Section4}.

\section{EFFECTIVE STIFFNESS FOR SATURATING NONLINEARITY}
We take force-deformation profile of stiffening springs
\begin{eqnarray}
\text{F}(\mathrm{\Delta}x) =(1+\gamma) k \mathrm{\Delta}x - 2 \gamma k \frac{\erf(\nu \mathrm{\Delta} x)}{c \nu}  \nonumber
\end{eqnarray}

\noindent and expand in power series, such that
\begin{IEEEeqnarray}{rCl}
\text{F}(\mathrm{\Delta}x)&=&(1-\gamma)k \mathrm{\Delta}x \nonumber \\
&&+\> \frac{2 \gamma k}{\nu} \bigg [ \frac{(\nu \mathrm{\Delta}x)^3}{3} - \frac{(\nu \mathrm{\Delta}x)^5}{5.2!} + \frac{(\nu \mathrm{\Delta}x)^7}{7.3!} - ...\bigg ].  \nonumber
\end{IEEEeqnarray}

\noindent Thus we write a general form
\begin{equation}
\text{F}(\mathrm{\Delta}x) =(1-\gamma)k \mathrm{\Delta}x + k_3 (\mathrm{\Delta}x)^3 - k_5 (\mathrm{\Delta}x)^5 +  ... \, , \nonumber  
\end{equation}

\noindent where the coefficient of $(\mathrm{\Delta}x)^n$ is defined as
\begin{eqnarray}
k_n =  2 \gamma k \bigg [  \frac{\nu^{n-1}}{n \big (\frac{n-1}{2} \big )!}   \bigg ], \; \; n = 3, 5, 7, ...  \nonumber
\end{eqnarray}

\noindent We also define a non-dimensional parameter to indicate the extent of nonlinearity
\begin{eqnarray}
\mathrm{\Gamma}_n = a^{n-1} \frac{k_n}{k}  = 2 \gamma  \bigg [ \frac{(a \nu)^{n-1}}{n \big (\frac{n-1}{2} \big )!}   \bigg ], \; \; n = 3, 5, 7, ... 
\end{eqnarray}

 We now write the equations of motion in non-dimensional form and employ similar procedure as was done in Appendix A. Thus we obtain the lower band-edge frequency
\begin{IEEEeqnarray}{rCl}
\mathrm{\Omega}_l^2 &=& 2 (1-\gamma) + 6  \mathrm{\Gamma}_3 A^2 - 20  \mathrm{\Gamma}_5 A^4+ 70  \mathrm{\Gamma}_7 A^6 \nonumber \\
&&-\> 252 \mathrm{\Gamma}_9 A^8+ 924\mathrm{\Gamma}_{11} A^{10} -  ...  \nonumber \\
&=& 2 (1-\gamma)  + 2\sum_{n=3,5,..} (-1)^{\frac{n+1}{2}} n C_{\frac{n-1}{2}}   \mathrm{\Gamma}_n  A^{n-1},  
\end{IEEEeqnarray}

\noindent where the Catalan number is defined as
\begin{eqnarray}
C_m=\frac{2m!}{(m+1)! m!}. \nonumber
\end{eqnarray}

\noindent Similarly, we can calculate the upper band-edge frequency
\begin{eqnarray}
\mathrm{\Omega}_u^2 = 2 (1+\gamma)  - 2\sum_{n=3,5,..} (-1)^{\frac{n+1}{2}} n C_{\frac{n-1}{2}}  \mathrm{\Gamma}_n  A^{n-1}. 
\end{eqnarray}

\noindent We, therefore, obtain effective stiffnesses for the springs in terms of strain amplitude $\mathrm{\Delta}A=2 A$
\begin{IEEEeqnarray}{rCl}
\text{K}_{\textrm{stiffening}}(\mathrm{\Delta}A) &=&\frac{\mathrm{\Omega}_l^2}{2} \nonumber \\ 
&=&(1-\gamma)  \nonumber \\
&&+\> \sum_{n=3,5,..} (-1)^{\frac{n+1}{2}} n C_{\frac{n-1}{2}}  \mathrm{\Gamma}_n   \bigg (\frac{\mathrm{\Delta}A}{2} \bigg )^{n-1}, \nonumber \\ 
\text{K}_{\textrm{softening}}(\mathrm{\Delta}A) &=&\frac{\mathrm{\Omega}_u^2}{2} \nonumber \\
&=&(1+\gamma) \nonumber  \\ 
&& -\> \sum_{n=3,5,..} (-1)^{\frac{n+1}{2}} n C_{\frac{n-1}{2}}  \mathrm{\Gamma}_n   \bigg (\frac{\mathrm{\Delta}A}{2} \bigg )^{n-1} . \nonumber
\end{IEEEeqnarray}

With this, we see how nonlinearity changes the effective stiffness of the springs. Increasing strain amplitude would make effective stiffnesses come closer to each other, and then go apart and saturate. The point (strain) where these stiffnesses would be equal, indicating the closure of the bandgap, is determined by a graphical method.  This strain amplitude explains the plateau in the edge mode profile shown for this case.

\bibliographystyle{apsrev4-1}

\bibliography{references.bib}

\end{document}